\documentclass[floatfix,superscriptaddress,numerical,amsmath,amssymb, pre, aps,onecolumn,showpacs,longbibliography]{revtex4-2}
\usepackage{graphicx}
\usepackage{caption}
\usepackage{paralist}
\usepackage{amsmath}
\usepackage{amssymb}
\usepackage{bm}
\usepackage{hyperref}
\usepackage{mathtools}

\usepackage{color}
\usepackage{verbatim}
\usepackage{comment,xcolor}

\usepackage[normalem]{ulem}

\begin{document}
\title{Polarization  and Consensus in a Voter Model under Time-Fluctuating Influences}

\author{Mauro Mobilia}
\affiliation{Department of Applied Mathematics, School of Mathematics, University of
Leeds, Leeds LS2 9JT, U.K.}

\begin{abstract}
We study 
 the effect of time-fluctuating social influences
on the formation of polarization and consensus  in  a three-party community consisting of two types of voters  (``leftists'' and ``rightists'') holding extreme opinions, and moderate agents acting as ``centrists''. The former are incompatible and  do not interact, while centrists hold an intermediate opinion and can interact with extreme voters. When
a centrist and a leftist/rightist interact, they can become either both centrists or both leftists/rightists.
The population eventually  either reaches consensus with one of the three opinions, or a polarization state consisting of a frozen mixture of leftists and rightists.
As a main novelty, here agents interact subject to time-fluctuating external influences favouring in turn the spread of leftist and rightist opinions, or  the rise of centrism.
The
fate of the population is determined
under various scenarios, and 
it is shown
how the rate of change of external influences  can drastically  affect the polarization and consensus probabilities, as well as the mean time to reach the final state. 
\end{abstract}

\maketitle
\section{introduction}
The relevance of parsimonious individual-based models to describe social phenomena at
micro and macro levels
has a long history \cite{Schelling,Granovetter,Galam1}. In the last few decades, ``sociophysics'' has grown as a research field that aims at studying 
collective social behaviour, like the spread of opinions or  
the dynamics of cultural diversity, using models and methods 
 from statistical physics~\cite{Galam1,Galam2,Galam3,Galam4,Castellano-rev,Galam-book,Sen-book,Perc-2017,Schweitzer2018,SW-2019,Redner-2019,Szolnoki-2022}. Typical questions in sociophysics concern 
 the conditions under which consensus, or long-term opinion diversity,  
emerges in a population of agents (``voters'') whose  states (``opinions'') change as they interact. 

The voter model (VM)~\cite{Liggett}, closely related  to the Ising model~\cite{Glauber}, has been commonly
used to describe how consensus ensues from the interactions between neighbouring
voters. In fact, while the classical two-state VM is arguably the simplest and most popular model of
opinion dynamics,  it rests on a number of oversimplifying assumptions: voters are endowed with only two possible states, they are  blind to any external stimuli,  have  zero self-confidence and are all identical. In reality,  members of social communities respond differently to stimuli, as they can interact in groups~\cite{Granovetter,Asch,Milgram,Lattane}, and are usually  characterised by multiple attributes~\cite{Axelrod,Axelrod-book,Castellano2000,Klemm2003,McPherson1987}.
In light of this, many generalizations  of the VM have been proposed:
for instance, ``zealotry'' was introduced in various forms to endow voters with different levels of self-confidence~\cite{Mobilia2003,Mobilia2005,Mobilia2007,Mobilia2013,Mobilia2015,zealotother1,zealotother2,zealotother3,zealotother4,zealotother5,zealotother6,zealotother7}, while group-size influence is notably captured in the nonlinear $q$-voter model~\cite{qVM} and its  variants~\cite{Sznajd,Slanina,genSznajd,exitprobq2,exitprobq,Mellor2016,Mellor2017,vacillating,Galam6}. 
It has also been noted that in many cases only some of the attributes characterising  agents are actually  compatible for social interactions~\cite{Axelrod,McPherson1987,BoundedCompromise1,BoundedCompromise2,BoundedCompromise3,BoundedCompromise4,BoundedCompromise5}. It was thus suggested that agents whose opinions are too different would  not interact,  while voters holding close opinions can interact and attain a global consensus.
This motivated the study of multi-state VMs, like the constrained three-state voter model (3CVM)  of Refs.~\cite{VKR2003,VR2004,MM2011}, which is  a discrete version of the bounded-compromise model~\cite{BoundedCompromise1,BoundedCompromise2,BoundedCompromise3,BoundedCompromise4,BoundedCompromise5}.
In the 3CVM, incompatible ``leftist'' and ``rightist'' voters can only interact with ``centrists'', and the final outcome is either  consensus with one of the three parties, or a polarized state consisting of mixture of leftists and rightists. In the latter case, the population is stuck in a frozen state of ``polarization'' in which 
leftists and rightists hold  uncompromising opinions~\cite{Galam23}.

In addition to interactions among agents, external stimuli or influences, especially social media and news sources, play an increasingly important role  in shaping the social environment that in turn affects  the
collective social behaviour~\cite{Granovetter,Media1,Media2,Media3,Media4,Media5,Media6,Media7}. Yet, with some recent remarkable exceptions~\cite{Bhat1,Bhat2,Media23,NatureCom2019},  the role of social influences
is still rarely modelled in sociophysics. 
Actually, 
sources of news play both a crucial, yet complex and multifaceted, role in influencing public social behaviour. They are  typically characterised by opposing viewpoints, or agenda,  that often change over time and can in turn favour consensus or polarization. It is also worth noting that voter-like models 
subject to a time-periodic  external field
have recently been studied in the context of economic networks~\cite{Hyst2,Hyst3}.

How do the opinions of a social community evolve in  a volatile and time-fluctuating environment?
Inspired by this question, here we introduce a generalization of  the constrained three-state voter model~\cite{VKR2003,VR2004,MM2011} in the presence of binary time-fluctuating external influences. In the same vein as in population  dynamics~\cite{Kimura,Ewens,AMR2013,WMR2018,Bena2006,HL06,Ridolfi11,Shnerb2015,Hufton2016,Hidalgo2017,KEM1,KEM2,TWAM,TWAM2}, for the sake of simplicity, we assume that the media influences  endlessly switch from favouring the spread of polarization 
to promoting centrism, and vice versa.
The goal of this work is to
determine the fate of the population under various scenarios,  and in particular to
study how the time variation of the external influences 
affects the probability to reach polarization or a consensus, as well as  the mean time for the population to settle in its final state.

The plan  of the paper is as follows: the general formulation of the model is introduced in the next section. Section III is dedicated to  a thorough study of the polarization and consensus probabilities. Section IV focuses on the study of the mean exit time. 
Section V is dedicated to a discussion of the results and to the conclusions.
 Technical details, including useful results  in the absence of external influences, and possible generalizations and applications are discussed in three appendices.
 
\section{Three-state constrained voter model under binary time-fluctuating  influences}
\begin{figure}[t]
\begin{center}
\includegraphics[width=4.5in,clip=]{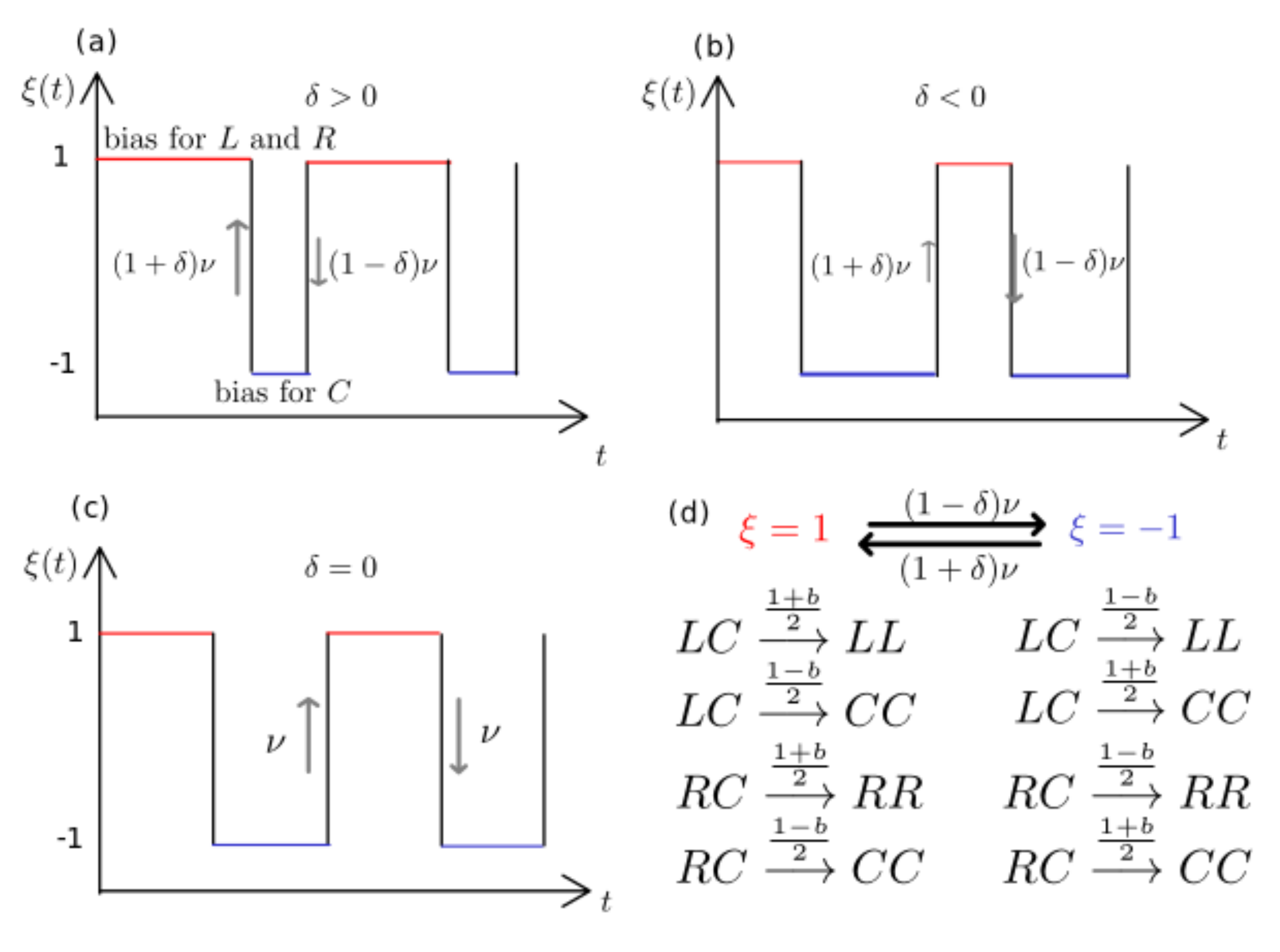}
\caption{
Illustration of the 3CVM under binary social switching $\xi \to -\xi$ at rate $(1-\delta \xi)\nu$ (where $\nu$ is the  average switching rate and $\delta$ the 
switching asymmetry).  ({\bf a}) $\xi(t)$ vs. time $t$ when $\delta>0$:
most time is spent in social state $\xi=1$ favouring  polarization. ({\bf b}) $\xi(t)$ vs.  $t$ when $\delta<0$: most time is spent in social state $\xi=-1$
favouring centrism. ({\bf c}) $\xi(t)$ vs.  $t$ when $\delta=0$: switching is symmetric  and  the same average time is spent in $\xi=\pm 1$. ({\bf d}) When  $\xi=1$ there is a bias favouring the spread of $L$ and $R$: $LC\to LL$ and $RC\to RR$  are the reactions with the highest rate, $(1+b)/2$ (where $b$ denotes the social influences bias). The social environment $\xi=-1$ favours the spread of centrism:  
the reactions  $LC\to CC$ and $RC\to CC$ have the highest rate, $(1+b)/2$, under $\xi=-1$. See text for details.
 In  ({\bf a})-({\bf c}), initially $\xi(0)=1$.}
  \label{Fig:Fig1_3VMswitch}
\end{center}
\end{figure}

We consider a well-mixed population of $N$ individuals consisting of  $N_R$ ``rightists'', or $R$-voters, $N_L$ ``leftists'', or  $L$-voters, and $N_C$ ``centrists'', or $C$-voters, with
$N=N_L+N_R+N_C$.
In the voter model language~\cite{Liggett,Castellano-rev,Galam-book,Sen-book,Redner-2019},  $R$ and $L$ represent extreme opinions,   while  $C$-voters hold an intermediate opinion.
In the three-state constrained voter model (3CVM)~\cite{VR2004,MM2011},
an agent selected at random tries to
 interact with one of its neighbours, that is any other randomly picked voter, at each
 microscopic update attempt. When the two agents hold the same opinion or if it is a pair of leftist-rightist ($LR$ or $RL$), nothing happens.  However, if the pair is a centrist $C$ and an 
 $L$ or an $R$ voter ($LC,~CL,~RC$ or $CR$), the initial agent adopts the opinion of the
neighbour with a probability that depends on the external influences whose effect is encoded into the  random variable $\xi(t)$ that fluctuates with the time $t$; see below and Fig.~\ref{Fig:Fig1_3VMswitch}.
Hence, while the interactions between $L$ and $R$ voters and centrists $C$ now change in time with $\xi$,  $L$ and $R$
remain incompatible and the system's final state is, as in the static 3CVM, either a consensus state or a state of polarization consisting of a frozen mixture of leftists and
rightists~\cite{VR2004,MM2011}; see below in this section. 

 %

 The main novel ingredient of this study is the modelling of time-fluctuating social influences  by means of the coloured dichotomous (telegraph) noise $\xi(t)\in \{-1,1\}$~\cite{Bena2006,HL06,Ridolfi11}; see Fig.~\ref{Fig:Fig1_3VMswitch}. The process $\xi(t)$ simply encodes all the 
complex effects of social environment and external influences in  the 
  endless random switching 
 between two  states: here, $\xi=1$ corresponds to external influences favouring $L$ and $R$ opinions (spread of polarization), whereas $\xi=-1$ favours
 centrism $C$ (compromise between  $L$ and $R$).
 Here, the dichotomous noise $\xi$ is always at \emph{stationarity},
 and switches from $\pm 1$ to $\mp 1$ according to
\begin{equation}
 \xi \longrightarrow -\xi, \label{switch}
\end{equation}
 with a rate $(1-\delta \xi)\nu$,
 where $\nu$ is the {\it average switching rate} (since $\nu=\frac{1}{2}[(1-\delta)\nu + (1+\delta)\nu]$)~\cite{TWAM}, and 
 $1<\delta<-1$ denotes the switching asymmetry. 
 Accordingly, the average time spent in state $\xi=\pm 1$
 before switching to $-\xi$ is $1/(1\mp\delta)\nu$ (symmetric switching occurs when $\delta=0$);  see Fig.~\ref{Fig:Fig1_3VMswitch}(a)-(c).
 At stationarity,
 $\xi=\pm 1$ with probability $(1\pm \delta)/2$~\cite{Bena2006,HL06,Ridolfi11}, and  its (ensemble-)average is 
 \begin{align}
 \label{delta}
  \langle \xi(t)\rangle&=\delta,
 \end{align}
 while its autocorrelation is $ \langle \xi(t)\xi(t')\rangle- \langle \xi(t)\rangle \langle \xi(t')\rangle
  =(1-\delta^2)e^{-2\nu|t-t'|}$.

\begin{figure}[t]
\begin{center}
\includegraphics[width=4.5in,clip=]{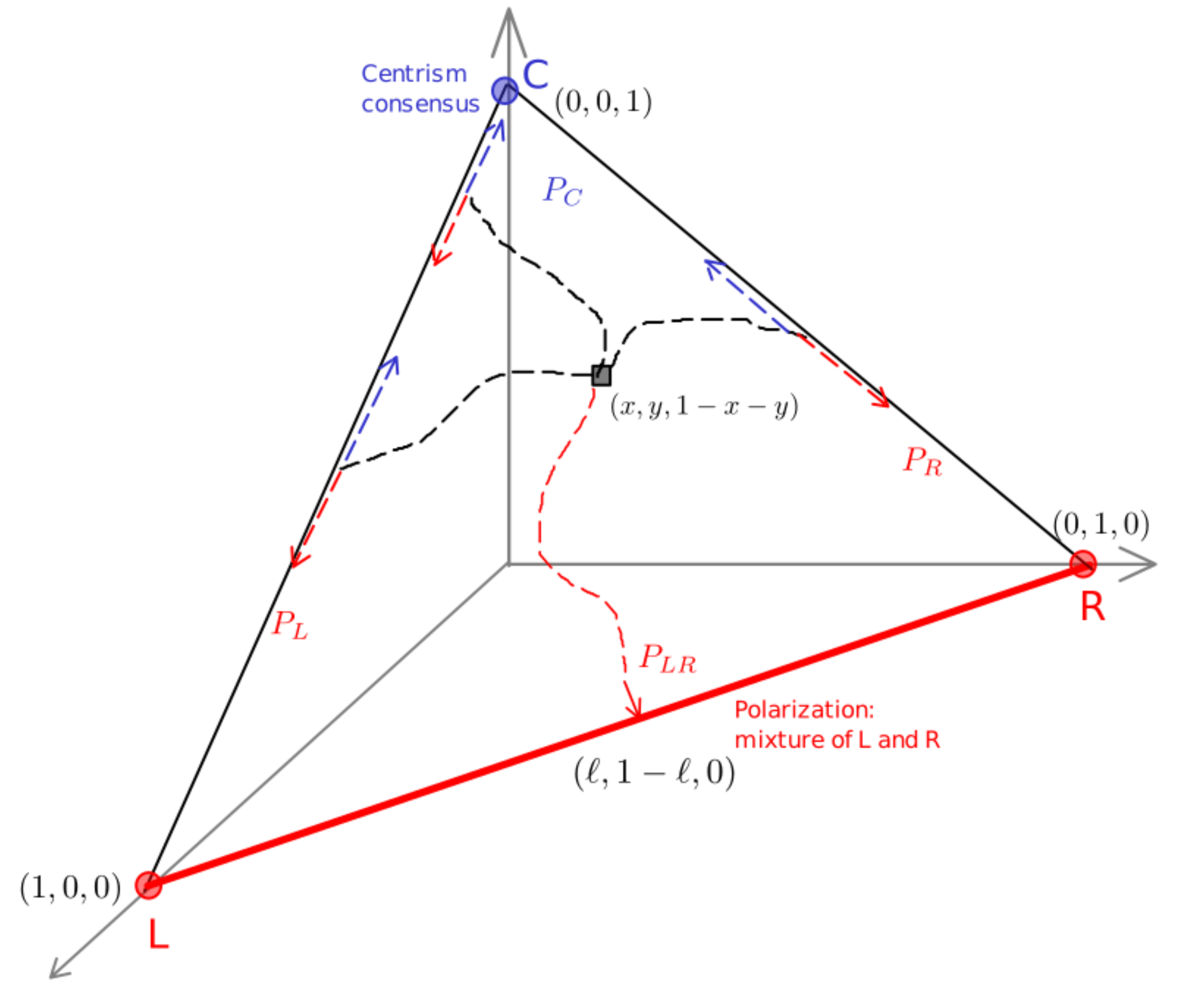}
\caption{
Illustration of the 3CVM dynamics in the simplex $\ell+r+c=(N_L+N_R+N_C)/N=1$. Circles show the consensus (absorbing) states all-$L$ $(1,0,0)$, all-$R$ $(0,1,0)$, and all-$C$ $(0,0,1)$, and the initial condition is $(x,y,z)$. The thick line indicates polarization state (pol-$LR$) consisting of a frozen mixture of $L$ and $R$ voters made up of a fraction $\ell=N_L/N$ of $L$-voters coexisting with an incompatible fraction $1-\ell$ of $R$-voters. Dashed lines are 
 typical trajectories: the dynamics ceases when the  line pol-$LR$ is reached (polarization with probability $P_{LR}$), or when there is consensus by reaching one of  absorbing states  all-$L$,  all-$R$
or  all-$C$ (with respective probabilities $P_L, P_R$ and $P_C$).
Once a trajectory reaches  the line $\ell=0$ or $r=0$, the evolution is restricted on $\ell=0$, $r=0$ until there is a consensus. As $\xi(t)$ varies, the dynamics favours in turn the spread of $L$ and $R$ ($\xi=1$) or that of $C$ ($\xi=-1$). See text for details.
}
  \label{Fig:Fig2_3VMswitch}
\end{center}
\end{figure}

The 3CVM switching dynamics is therefore defined
by the four reactions: $LC \to LL$ and $RC \to RR$, corresponding to the spread of 
extreme opinions at rate $(1+b\xi)/2$,
and $LC \to CC$ and $RC \to CC$,   with
 centrists replacing $L$ and $R$ voters  at rate $(1-b\xi)/2$.
Here, $0< b< 1$ denotes the social influences bias favouring polarization when $\xi=1$ and centrism in the social environment  $\xi=-1$. 
The 3CVM switching 
dynamics  can thus be schematically described by the following reactions occurring at each time increment:
\begin{eqnarray*}
  L  C &\longrightarrow& LL\quad {\rm  \ rate:} \ \frac{1+b \xi}{2}, \qquad  \   
  L  C \longrightarrow C  C \quad {\rm  \ rate:} \ \frac{1-b \xi}{2} \ 
  \nonumber \\
  R  C &\longrightarrow& R  R \quad {\rm  \ rate:} \ \frac{1+b\xi}{2}, \qquad  \  
  R  C \longrightarrow C  C  \quad {\rm  \ rate:} \ \frac{1-b \xi}{2}, \label{react}
\end{eqnarray*}
where $\xi$ and the rates  endlessly fluctuates according to \eqref{switch}. The dynamics of the switching 3CVM is therefore
the Markov chain defined by the transition rates \eqref{transrate}
and master equation \eqref{eq:ME}~\cite{Gardiner}; see Appendix \ref{AppA1}.
The 3CVM switching dynamics is characterised by 
three consensus/absorbing states    $N_R=N_C=0$ (all-$L$), $N_L=N_C=0$ (all-$R$), $N_L=N_R=0$, $N_C>0$ (all-$C$),
and by the polarization state  (pol-$LR$) where $N_L+N_R=N$, $N_C=0$; see Fig.~\ref{Fig:Fig2_3VMswitch}. As in the absence of external influences, the final state of the population is therefore guaranteed to be either one of the consensus/absorbing states or pol-$LR$~\cite{VR2004,MM2011}. 

 It is worth noting that the approach considered here bears some similarities with the two-party model of Refs.~\cite{Bhat1,Bhat2}. However there are also important differences: first, in the 3CVM, the polarization state corresponds to a frozen mixture (while it is an active state in Refs.~\cite{Bhat1,Bhat2}). Moreover, external influences  are here assumed to fluctuate endlessly in time, rather than by establishing a certain number of connections with voters.
 It is also interesting to notice that the effect of an exogenous time-varying influence has recently been studied in the context of  economic networks~\cite{Hyst2,Hyst3}. 
\ The ensuing dynamics derived also from 
 an Ising-like (voter-like) model subject to a time-dependent external field~\cite{Hyst2,Hyst3} has been shown to lead to rich dynamics characterized by hysteresis~\cite{Hyst1}, which is 
 not a phenomemon exhibited by the   3CVM switching dynamics. 
 This stems from the fact that Eqs.~(\ref{SDE}) differ from those governing the mean-field dynamics in Refs.~\cite{Hyst2,Hyst3} for not having any non-noisy linear terms, and for the 
the exogenous time dependence being stochastic 
 (via the multiplicative dichotomous noise $\xi(t)$) rather than periodic.


\section{Final state: polarization and consensus probabilities}
\label{sec:Final}
In its final state the population is either in the polarized state pol-$LR$, or in one of its three absorbing/consensus states (all-$L$, all-$R$ or all-$C$); see Fig.~\ref{Fig:Fig2_3VMswitch}. We denote by $P_{LR}$ the probability to end up in the polarized final state pol-$LR$. The probabilities to reach the absorbing/consensus states all-$C$, all-$L$, and all-$R$ are respectively denoted by $P_C,P_L$  and $P_R$. The density of voters of each type is 
$\ell\equiv N_L/N,~r\equiv N_R/N$ and $c\equiv N_C/N=1-\ell-r$, and initially the population consists of densities $x,~y,~z=1-x-y$ of $L, R$ and $C$ voters, respectively.

In the absence of environmental switching, the probabilities of ending in any of the absorbing/consensus or polarization states were  found to depend non-trivially on the parameter $s\equiv Nb$ and $(x,y,z)$~\cite{VR2004,MM2011}; see Appendix~\ref{AppB}. Here, we are interested in finding how $P_{LR}, P_C,P_L$  and $P_R$, as well as the final densities $(\ell,r,c)$  depend on $\nu$ under various scenarios.
 We consider the same initial density of $L$ and $R$ voters, i.e. $0<x=y=(1-z)/2<1/2$, which suffices for for 
 the
purposes of this study  and simplifies the analysis. (Only in Appendix~\ref{AppC}, an example with  $x\neq y$ is briefly considered.) By symmetry, $x=y$ implies $P_L(\nu)=P_R(\nu)$, i.e. the  probability of $L$ and $R$ consensus is the same. When it occurs, polarization consists of a fraction $1/2$ of $L$ and $R$ voters; see Fig.~\ref{Fig:Fig2_3VMswitch}. We thus have: $P_{LR}+P_C+P_L+P_R=P_{LR}+P_C+2P_L=1$, and therefore focus on studying $P_{LR}$ and $P_{C}$ as functions of $\nu$ for different values of $\delta$ and $z$, treated as parameters, from which we obtain also the final densities: $(\ell,~r,~c)=(\ell,~\ell,~1-2\ell)=((1-P_C)/2,~(1-P_C)/2,~P_C)$; see Appendix~\ref{AppA2}.

\subsection{Final state in the regimes $\nu \to 0$ and $\nu \to \infty$}
\label{sec:Prob}
The  polarization and consensus probabilities
can be computed analytically in the regimes $\nu \to 0$ and $\nu \to \infty$. For this, we take advantage of the results obtained in the {\it absence of external influences}  for 
the polarization probability ${\cal P}_{LR}$,
and the probabilities of  $C, L$ and $R$ consensus, respectively denoted here by ${\cal P}_{C}$, ${\cal P}_{L}$
and ${\cal P}_{R}$. In the 
absence of external  influences,
these quantities have been obtained   in Refs.~\cite{VR2004,MM2011}
and are summarized in Appendix~\ref{AppB1}.

\subsubsection{Polarization and consensus probabilities in the regime $\nu\to 0$}
When $\nu\to 0$, we can assume that there are no  switches before attaining polarization or consensus. In this case, $\xi$ is a quenched random variable,
and the kinetics is the superposition, with probability  $(1\pm\delta)/2$, of the 3CVM dynamics in the stationary environment $\xi=\pm 1$. 
As a result, the polarization probability when $\nu\to 0$, $P_{LR}^0$,   is the superposition of  ${\cal P}_{LR}(\pm s,z)$, obtained in a static external state
$\xi=\pm 1$, with probability $(1\pm\delta)/2$:  
\begin{align}
\label{PLR0}
P_{LR}^0=
 \left(\frac{1+\delta}{2}\right){\cal P}_{LR}(s,z)+\left(\frac{1-\delta}{2}\right){\cal P}_{LR}(-s,z),
\end{align}
which is  readily obtained from \eqref{Pab-stat} or \eqref{Pstat_nosa}.
 Similarly, the consensus probabilities when $\nu\to 0$,  are  obtained from \eqref{Pab-stat}, with \eqref{Pc-stat} and \eqref{PA_stat}, and ${\cal P}_{L}(s,z)=(1-{\cal P}_{LR}(s,z)-{\cal P}_{C}(s,z))/2$:
\begin{align}
\label{PCL0}
P_{C,L}^0=
 \left(\frac{1+\delta}{2}\right){\cal P}_{C,L}(s,z)+\left(\frac{1-\delta}{2}\right){\cal P}_{C,L}(-s,z).
\end{align}

With \eqref{lr_stat} and \eqref{PCL0}, we  obtain the final density $\ell_{0}=r_0$ of $L$ and $R$ voters when $\nu\to 0$:
\begin{align}
\label{l0}
 \ell_{0}=r_0= \frac{1-P_{C}^0}{2}.
\end{align}

\subsubsection{Polarization and consensus probabilities in the regime $\nu\to \infty$}

When $\nu\to \infty$, so many switches occur before polarization or consensus that $\xi$ self-averages~\cite{HL06,KEM1,KEM2,WMR2018,TWAM,TWAM2}. In this case, $\xi$ is an annealed random variable that can be replaced by its average: $\xi\to \langle \xi\rangle=\delta$.
The switching dynamics of the 3CVM with $\nu\to \infty$ is thus the same as in Ref.~\cite{MM2011}, with $s\to s\langle \xi\rangle=s\delta$.
 In the limit $\nu\to \infty$, the polarization probability, 
$P_{LR}^{\infty}$,  is therefore
obtained from  \eqref{Pab-stat} or \eqref{Pstat_nosa}
according to  
\begin{align}
\label{PLRinf}
P_{LR}^\infty=
 {\cal P}_{LR}(s\delta,z).
\end{align}
Similarly,  the consensus probabilities under high $\nu$, $P_{C,L}^\infty$, are obtained from  \eqref{Pc-stat} and \eqref{Pstat_nosa}:
\begin{align}
\label{PCLinf}
P_{C,L}^\infty=
 {\cal P}_{C,L}(s\delta,z).
\end{align}

With \eqref{lr_stat} and \eqref{PCLinf},
the final density $\ell_{\infty}=r_{\infty}$ of $L$ and $R$ voters in the regime $\nu\to \infty$:
\begin{align}
\label{linf}
 \ell_{\infty} =r_{\infty}
 =\frac{1- P_{C}^{\infty}}{2}.
\end{align}

Since ${\cal P}_{LR}\approx 1$ when $s\equiv Nb\gg 1$
and ${\cal P}_{C}\approx 1$ when $s<0$ and $|s|\gg1$~\cite{MM2011} (see Appendix~\ref{AppB1}),  the focus here is
on the regime where the small influences bias
affects a large but finite number of voters, i.e.  $b\ll 1$ and $s\gg 1$,  with $s\delta={\cal O}(1)$. This allows us to highlight the effect of the external influences.

\subsection{Polarization and consensus probabilities when $\delta>0$}
\begin{figure}[t!]
\begin{center}
\includegraphics[width=4.5in,clip=]{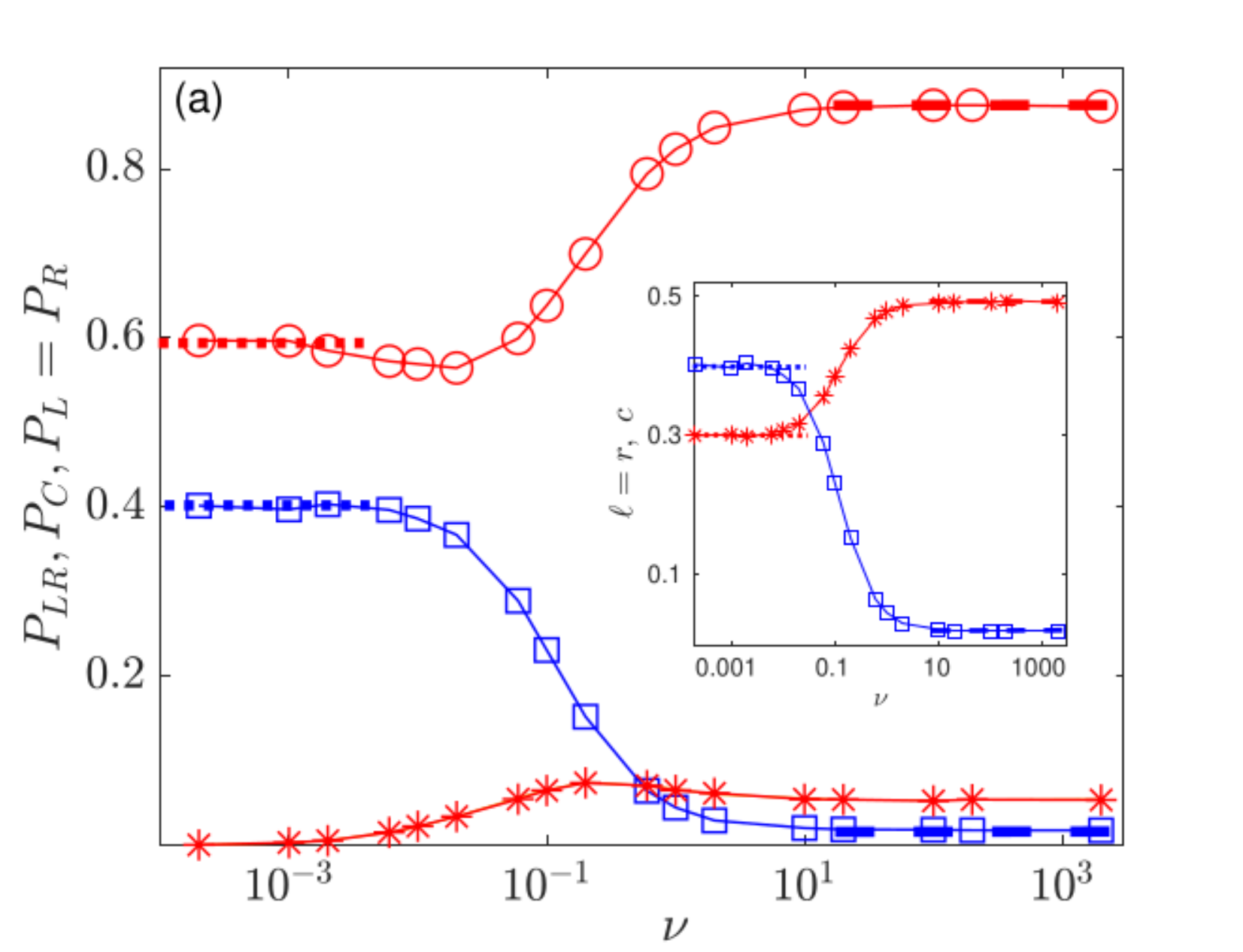}
\includegraphics[width=4.5in,clip=]{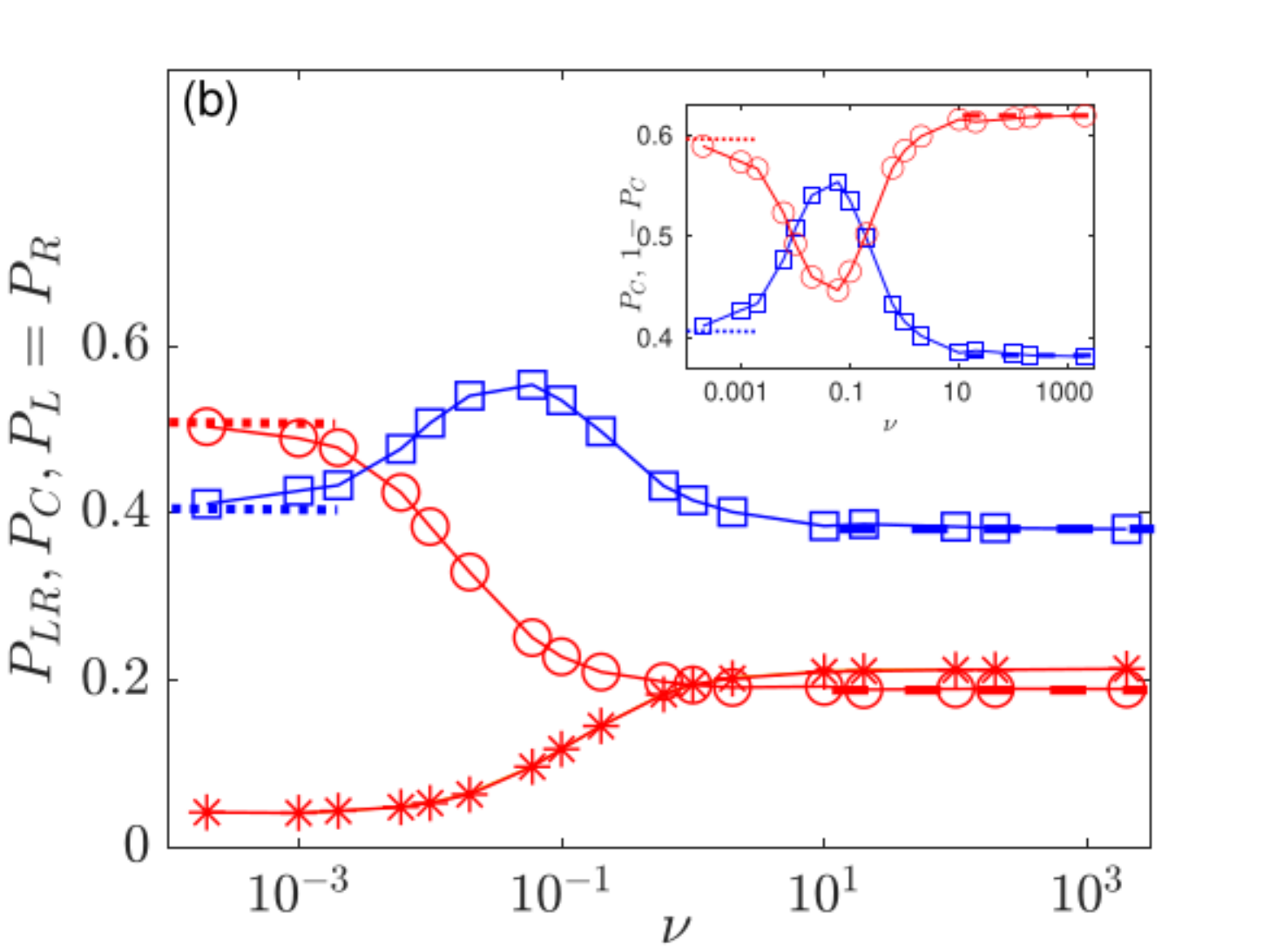}
\caption{ $P_{LR}$ (circles), $P_C$ (squares), $P_L=P_R$ (stars)
vs. $\nu$ for $\delta=0.2$ and different initial conditions.
Here and in the other figures symbols are from simulations (averaged over $10^5$ samples).
Thick dashed horizontal lines are eyeguides showing $P_{LR}^{\infty}$ (red), $P_{C}^{\infty}$ (blue), and thick dotted lines are eyeguides showing 
$P_{LR}^{0}$ (red), $P_{C}^{0}$ (blue) given by \eqref{PLRinf},\eqref{PCLinf},\eqref{PLR0}, and \eqref{PCL0}. 
({\bf a}) Initial condition: $x=y=0.25$ and $z=0.5$. 
Inset: final densities $\ell=r$ 
($L/R$-voters, stars) and $c$ (centrists, squares) vs. $\nu$.  Horizontal  lines show  $\ell_{\infty}=r_{\infty}$ (dashed, red) and  $c_{\infty}=1-2\ell_{\infty}$ (dashed, blue)
from
\eqref{linf},  $\ell_{0}$ (dotted, red) and  $c_{0}=1-2\ell_{0}$ (dotted, blue) from \eqref{l0}.
({\bf b}) Initial condition: $x=y=0.06$ and $z=0.88$.
Inset:
$P_C$ (squares) and $1-P_C$ (diamonds) vs. $\nu$: centrism is the majority opinion in the range of intermediate switching rate, $\nu\sim b=0.1$, where $P_C(\nu)>1/2$.
The dashed lines show
 $1-P_{C}^{\infty}$ (red), $P_C^{\infty}$ (blue), and the dotted lines show $1-P_{C}^{0}$ (red) and  $P_C^{0}$ (blue).
In all panels and insets: $(N,b,s)=(200, 0.1, 20)$.  See text for more details.
}
  \label{Fig:Fig3_3VMswitch}
\end{center}
\end{figure} 
When $\delta>0$, most of the time is spent in the external state
$\xi=1$, where influences  favour
 polarization (pol-$LR$); see Figs.~\ref{Fig:Fig1_3VMswitch} and \ref{Fig:Fig2_3VMswitch}.

In fact, as shown in Fig.~\ref{Fig:Fig3_3VMswitch}(a), when $z$ is not too close to $1$, $P_{LR}>P_{C}$ for all values of  $\nu$. In this case, $P_{LR}$ increases with $\nu$ over a large range of values (for $\nu\gtrsim b$), with $P_{LR}\approx P_{LR}^0\approx(1+\delta)/2$ when $\nu \ll b$ and $P_{LR}\approx P_{LR}^{\infty}$ when $\nu \gg b$, whereas $P_{C}$ is a decreasing function of $\nu$, with $P_C\approx P_{C}^0\approx(1-\delta)/2$ when $\nu \ll b$ 
and $P_{C}\approx P_{C}^{\infty}\ll 1$ when $\nu \gg b$; see Fig.~\ref{Fig:Fig3_3VMswitch}(a). %
The fact that 
$P_{LR}$ and $P_{C}$ vary little with $\nu$ and are very close 
to  $P_{LR}^{0,\infty}$ and $P_{C}^{0,\infty}$
indicates that the analytical predictions for $P_{LR}^{0,\infty}$ and $P_{C}^{0,\infty}$  not only apply to the limits $\nu\to 0,\infty$,
but are also valid approximations of $P_{LR}(\nu)$ and $P_{C}(\nu)$ in the regimes of low and high switching rate (see also Sec.~\ref{sec:MET} below). 
We can also notice in Fig.~\ref{Fig:Fig3_3VMswitch}(a), that $P_{LR}$ exhibits a weak non-monotonic behaviour, with a ``dip'' for $\nu\sim b$. 
In this setting, the final fraction of 
$L$  and $R$ voters, given by $\ell=r=(1-P_C)/2$ (see Appendix~\ref{AppA2}), is an increasing function of $\nu$, while the final density of centrists, $c=1-2\ell=P_C$,
decreases with $\nu$; see inset in Fig.~\ref{Fig:Fig3_3VMswitch}(a).

 When there is a small initial fraction of $L$ and $R$ voters, with $z$  close to $1$, centrism can prevail over  a large range of values of $\nu$: $P_{C}>P_{LR}$
 for $\nu\gtrsim b$; see  Fig.~\ref{Fig:Fig3_3VMswitch}(b). The probability $P_{LR}$ thus decreases with $\nu$, while  $P_{C}$ can have a non-monotonic behaviour as in Fig.~\ref{Fig:Fig3_3VMswitch}(b) where it exhibits a  ``bump'' for $\nu\sim b$. In this case,
 centrists hold the majority opinion
 over an intermediate range of $\nu$:
 $P_{C}>P_{LR}+P_L+P_R=1-P_C$, i.e. $P_C>1/2$ when $\nu={\cal O}(b)$; see inset in Fig.~\ref{Fig:Fig3_3VMswitch}(b).

 In Fig.~\ref{Fig:Fig3_3VMswitch}, the predictions 
 \eqref{PLR0}, \eqref{PCL0} and \eqref{PLRinf}, \eqref{PCLinf} for $P_{LR/C}^{0}$ and $P_{LR/C}^{\infty}$
are  in good agreement with simulation data when 
$\nu\ll b$ and $\nu\gg b$, respectively. %
The results reported in  Fig.~\ref{Fig:Fig3_3VMswitch} show that when $\delta>0$
the main effects of the switching  external influences is to tame the bias favouring  polarization. $P_{LR}$ and $P_C$ can thus increase or decrease with $\nu$, and  even have a local extremum
at a nontrivial switching rate. We  explain this  by solving, at fixed $\delta$,    $P_{LR}^{\infty}=P_{LR}^{0}$ and $P_{C}^{\infty}=P_{C}^{0}$ for $z_{LR}$ and $z_C$ with  \eqref{PLR0}, \eqref{PCL0}, \eqref{PLRinf}, \eqref{PCLinf}:
\begin{align}
\label{zLRzC}
\hspace{-5mm}
{\cal P}_{LR}(s\delta,z_{LR})&=\left(\frac{1+\delta}{2}\right){\cal P}_{LR}(s,z_{LR})+\left(\frac{1-\delta}{2}\right){\cal P}_{LR}(-s,z_{LR}), \nonumber\\
{\cal P}_{C}(s\delta,z_C)&=\left(\frac{1+\delta}{2}\right){\cal P}_{C}(s,z_{C})+\left(\frac{1-\delta}{2}\right){\cal P}_{C}(-s,z_{C}).
\end{align}
Using  \eqref{Pab-stat} and \eqref{Pc-stat}, these
equations are solved for
 $z_{LR}$ and $z_{C}$. When $z<z_{LR}$, $P_{LR}^{\infty}>P_{LR}^0$ while $P_{LR}^{\infty}<P_{LR}^0$  when $z>z_{LR}$. Similarly, $P_{C}^{\infty}>P_{C}^0$ when $z>z_C$ and $P_{C}^{\infty}<P_{C}^0$ for $z<z_C$.
In the examples of Fig.~\ref{Fig:Fig3_3VMswitch}, $z_{LR}\approx 0.708$ and $z_{C}\approx 0.887$, and therefore $z<z_{LR,C}$ in Fig.~\ref{Fig:Fig3_3VMswitch}(a) and 
 $z_{LR}<z<z_{C}$ in Fig.~\ref{Fig:Fig3_3VMswitch}(b), which agrees with $P_{LR}^{\infty}>P_{LR}^0$ and $P_{C}^{\infty}<P_{C}^0$ reported in Fig.~\ref{Fig:Fig3_3VMswitch}(a), and with $P_{LR,C}^{\infty}<P_{LR,C}^0$ in Fig.~\ref{Fig:Fig3_3VMswitch}(b).

The extrema of $P_{LR}$ and $P_C$ in Fig.~\ref{Fig:Fig3_3VMswitch} at $\nu\sim b$ can be explained heuristically 
(and similarly for those in Fig.~\ref{Fig:Fig4_3VMswitch} while there are no extrema  in Fig.~\ref{Fig:Fig5_3VMswitch}; see below). 
In fact, when $\nu\to 0$, the population remains in the initial external state $\xi(0)$ until polarization or consensus occurs. When $s\gg 1$ and $\nu\ll 1$, polarization occurs with a probability close to $1$ if $\xi(0)=1$ and close to $0$ otherwise, and we have $P_{LR}(\nu\ll b)\approx (1+\delta)/2$. We  argue  that  
$P_{LR}(\nu)$ decreases with $\nu$ when this rate is raised from $\nu\to0$ to $\nu\lesssim b\ll 1$. For the sake of argument, we focus on the switching rate $\nu_*$  for which there is one external switch before reaching the final state. As discussed in Sec.~\ref{sec:METgen}, we have $\nu_*\lesssim b$ (see the inset in Fig.~\ref{Fig:Fig6_3VMswitch}(b)), with 
$\xi(t>1/\nu^*)=-\xi(0)$, and the population settles in its final state after a time $T\sim (\ln{N})/b$.  Hence, if $\xi(0)=1$, the final state is polarization 
if {\rm pol}-LR is reached in a time  $t\lesssim 1/\nu_*$, when  still $\xi(t)=1$.
This occurs approximately with a probability $1/(T\nu_*)$. On the other hand,  if $\xi(0)=-1$, the final state is polarization if {\rm pol}-LR is reached after the external switch, 
when $\xi(t)=1$, i.e. for  $t\gtrsim 1/\nu_*$, and this occurs with an approximate  probability $1-1/(T\nu_*)$. Hence, we estimate \[P_{LR}(\nu_*)\approx \left(\frac{1+\delta}{2}\right)\frac{1}{T\nu_*}
+
\left(\frac{1-\delta}{2}\right)\left(1-\frac{1}{T\nu_*}\right)=\frac{1}{2}+\delta\left(\frac{1}{T\nu_*}-\frac{1}{2}\right)<P_{LR}^0,\]
which explains that $P_{LR}(\nu)$ decreases with $\nu$ when $\nu\lesssim b$. When $\nu$ is increased further above $b$, $P_{LR}(\nu)$ increases with $\nu$ and approaches $P_{LR}^{\infty}>P_{LR}^0$. This  explains qualitatively the non-monotonic behaviour of
$P_{LR}(\nu)$ in Fig.~\ref{Fig:Fig3_3VMswitch}(a), with a dip   
at 
$\nu_*\approx 0.02$, and $P_{LR}(\nu_*)\approx 0.565<P_{LR}^0\approx 0.6<P_L^{\infty}\approx 0.876$, while the rough estimate,
with $T\approx 63$ (see Fig.~\ref{Fig:Fig6_3VMswitch}(a)),
gives $P_{LR}(\nu_*)\approx 0.56$.
 Similarly, in Fig.~\ref{Fig:Fig3_3VMswitch}(b), $ P_{C}^{\infty}\approx P_{C}^{0}<P_{C}(\nu_*)$, which results in a ``bump''  in  $P_{C}(\nu)$ at some $\nu_*\lesssim b$.

\subsection{Polarization and consensus probabilities when $\delta<0$}
\begin{figure}[h!]
\begin{center}
\includegraphics[width=4.5in,clip=]{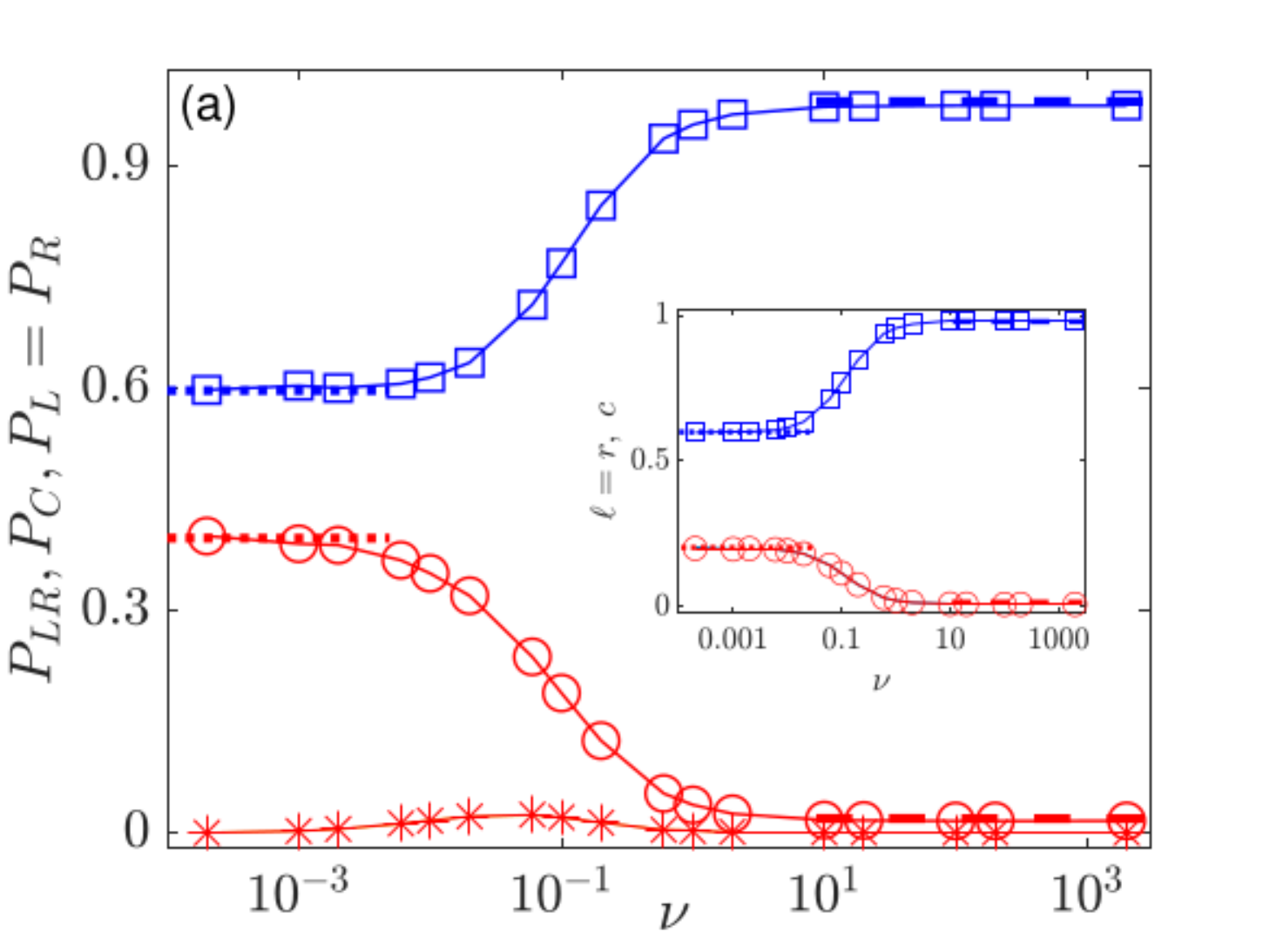}
\includegraphics[width=4.5in,clip=]{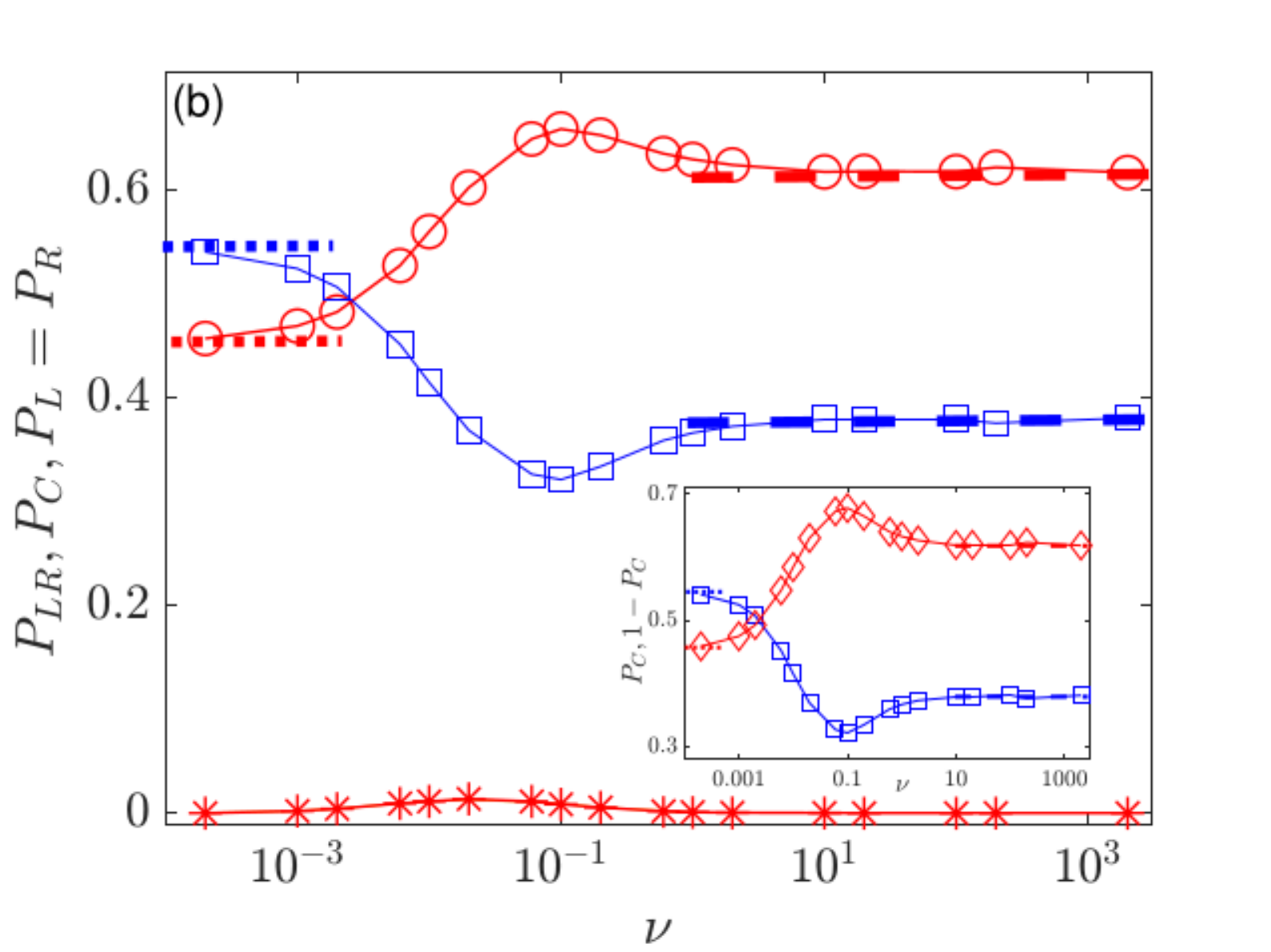}
\caption{ $P_{LR}$ (circles), $P_C$ (squares), $P_L=P_R$ (stars)
vs. $\nu$ for $\delta=-0.2$, and different initial conditions. 
Symbols are results from simulations (averaged over $10^5$ samples).
Thick dashed and dotted lines are as in Fig.~\ref{Fig:Fig3_3VMswitch}.
({\bf a}) Initial condition: $x=y=0.25$ and $z=0.5$. 
Inset: final densities $\ell=r$ 
($L/R$-voters, stars) and $c$ (centrists, squares) vs. $\nu$.  
({\bf b}) Initial condition: $x=y=0.47$ and $z=0.06$. Inset: 
$P_C$ (squares) and $1-P_C$ (diamonds) vs. $\nu$: centrists hold the majority opinion in the range of low switching rate ($\nu\ll b$) where $P_C(\nu)>1/2$.
In all panels and insets: $(N,b,s,\delta)=(200, 0.1, 20,-0.2)$. See text for more details.
}
  \label{Fig:Fig4_3VMswitch}
\end{center}
\end{figure}

When $\delta<0$, most time is spent in the external state 
$\xi=-1$, where  influences  favour
centrist consensus. Hence, when $s=Nb\gg 1$ and $z$ is not too close to $0$, 
centrism prevails over the other opinions:
$P_{C}>P_{LR}\gg P_{L,R}$ for all values of  $\nu$;
see Fig.~\ref{Fig:Fig4_3VMswitch}(a). In this case, we find: $P_{C}\approx P_{C}^0\approx(1+|\delta|)/2>P_{LR}\approx P_{LR}^0\approx (1-|\delta|)/2$ when $\nu \ll 1$, while 
$P_{C}\approx P_{C}^{\infty}\approx 1$ and $P_{LR} \approx P_{L,R}\approx P_{LR}^{\infty}\approx 0$ when $\nu \gg 1$. 
 In Fig.~\ref{Fig:Fig4_3VMswitch}(a),  $P_{C}$ increases with $\nu$ from $(1+|\delta|)/2$ to $1$, while,   $P_{LR}$ decreases from $(1-|\delta|)/2$ to $0$ as  $\nu$ is raised. This results in a final state consisting of a majority of $C$ voters, whose final density $c$ increases from $(1+|\delta|)/2$ to $1$, and a minority of $L$ and $R$ voters, whose final density $\ell=r$ decreases with $\nu$ from $(1-|\delta|)/4$ to $0$; see the inset in Fig.~\ref{Fig:Fig4_3VMswitch}(a). However, while all-$C$ is the most likely final state, there is a finite probability of polarization at low and intermediate switching rate.

 When $s\gg 1$ and the initial population consists mainly of $L$ and $R$ voters ($z\ll 1$), polarization  is the most likely final state,
 with $P_{LR}>P_C>P_{L,R}$  for $\nu\gg b$;
 see  Fig.~\ref{Fig:Fig4_3VMswitch}(b). Centrists thus generally hold the minority opinion when $\nu\gg b$, while $C$ is the majority opinion 
 ($P_C>1/2$) only under low switching rate; see inset in Fig.~\ref{Fig:Fig4_3VMswitch}(b). 
 In the limiting regimes $\nu\ll b$ and $\nu \gg b$, $P_{LR}$ and $P_C$  respectively approach the  values of $P_{LR}^{0,\infty}, P_C^{0,\infty}$. 
 Here again,  Equation \eqref{zLRzC} can be used 
 to determine the initial density $z_{LR}$,
 such that $P_{LR}^{\infty}>P_{LR}^{0}$ if $z<z_{LR}$ and $P_{LR}^{\infty}\leq P_{LR}^{0}$ otherwise, and  $z_{C}$ such that $P_{C}^{\infty}>P_{C}^{0}$ if $z>z_{C}$ and $P_{C}^{\infty}\leq P_{C}^{0}$ otherwise. In the example of Fig.~\ref{Fig:Fig4_3VMswitch}, $z_{LR}\approx z_{C}\approx 0.112$. As for $\delta>0$, in the regime of intermediate switching, as for $\delta>0$,
$P_{LR}$ and $P_{L,R}$
 can exhibit bumps and $P_C$ a dip; see Fig.~\ref{Fig:Fig4_3VMswitch}(b) 
 where  $P_{LR}, P_{L,R}$ and $P_{C}$ have   modest extrema around $\nu\sim b$.

\subsection{Polarization and consensus probabilities under symmetric switching ($\delta=0$)}
When $\delta=0$, social switching is symmetric, and the same average amount of time is spent in $\xi=\pm1$. External influences thus favour centrism ($\xi=-1$) and polarization ($\xi=1$) in turn (see Fig.~\ref{Fig:Fig1_3VMswitch}c), and all realizations start in either of the external states $\xi=\pm 1$ with the same probability $1/2$.

\begin{figure}[h!]
\begin{center}
\includegraphics[width=4.5in,clip=]{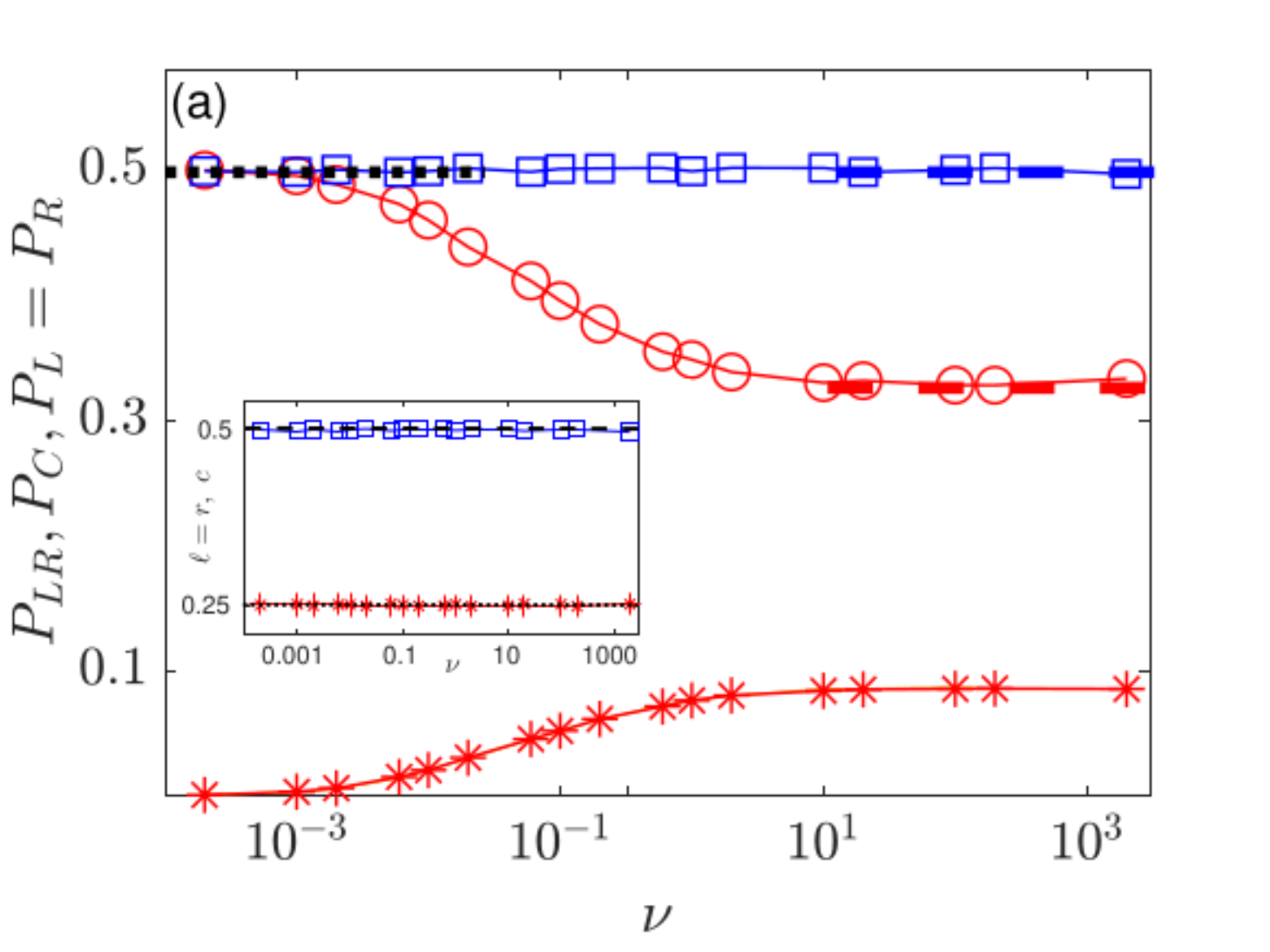}
\includegraphics[width=4.5in,clip=]{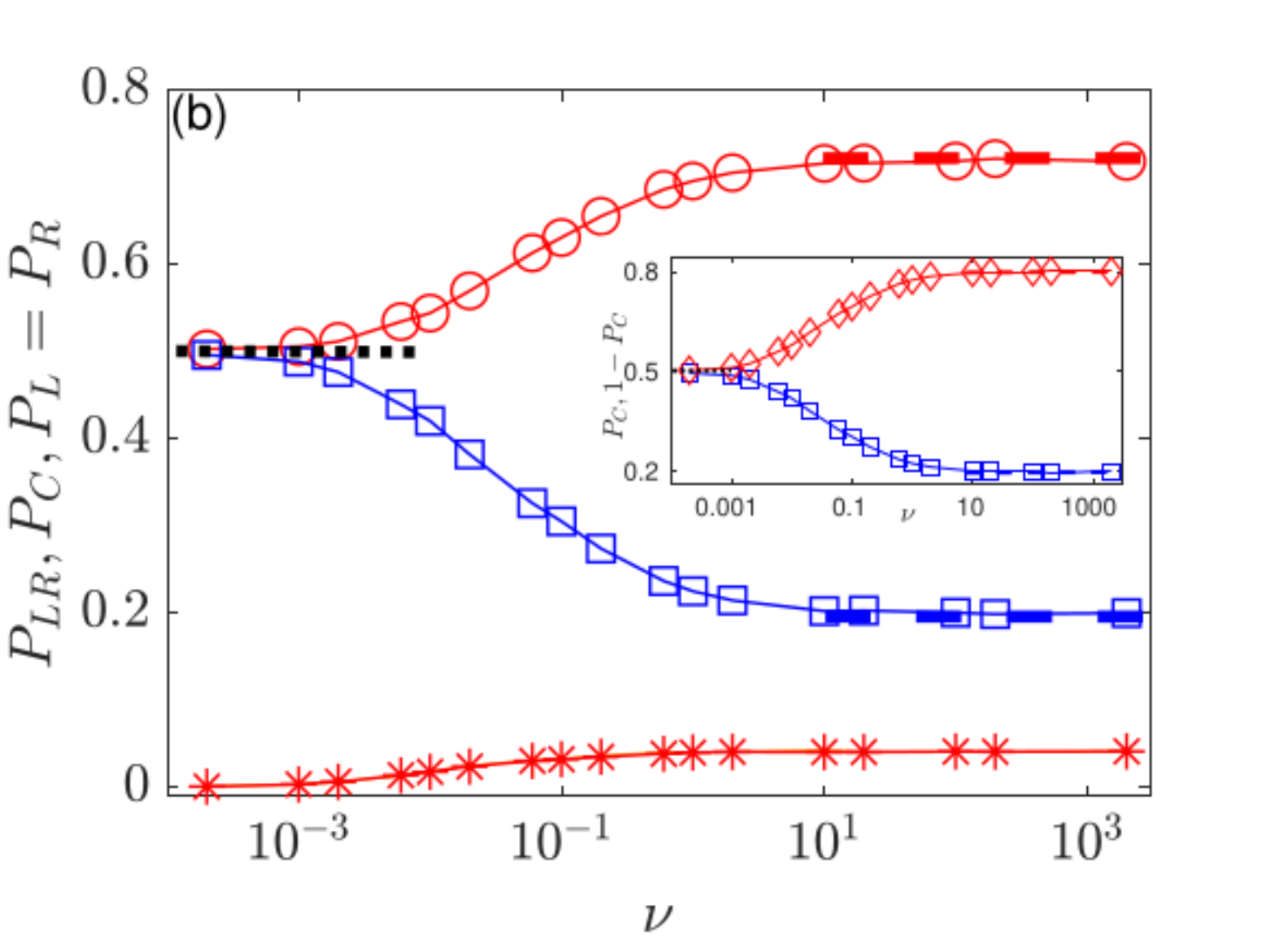}
\caption{$P_{LR}$ (circles), $P_C$ (squares), $P_L=P_R$ (stars)
vs. $\nu$ for $\delta=0$, and different initial conditions
under symmetric switching. Symbols are results from simulations (averaged over $10^5$ samples).
Thick dashed and dotted lines are as in Fig.~\ref{Fig:Fig3_3VMswitch}.
({\bf a}) Initial condition: $x=y=0.25$ and $z=0.5$. 
Inset: final densities $\ell=r$ 
($L/R$-voters, stars) and $c$ (centrists, squares) vs. $\nu$. In the special case $s\gg 1, ~\delta=0,~z=2x=2y$, $P_C(\nu)=c\approx 0.5$.  
({\bf b}) Initial condition: $x=y=0.4$ and $z=0.2$.
Inset: 
$P_C$ (squares) and $1-P_C$ (diamonds) vs. $\nu$: under $\delta=0$, centrists hold the minority opinion when $z<1/2$.
In all panels and insets, other parameters are: $(N,b,s,\delta)=(200,0.1,20,0)$. See text for more details.}
  \label{Fig:Fig5_3VMswitch}
\end{center}
\end{figure}

In the regime where $\nu\ll b$, the final state is reached with a probability close to $1/2$ from either  external state $\xi=\pm 1$.
When $\xi=1$, polarization is almost certain, whereas there is centrist consensus with probability $1$ when $\xi=-1$. Hence, $P_{LR}\approx P_C\approx P_{LR}^0=P_C^0=1/2$ and $P_L=P_R\approx 0$ when  $\nu\ll b$.
Under low switching rate, $\nu< b$, a small number switches can occur prior reaching the final state, but switching  being symmetric, the net effect mostly cancels out and $P_{LR}=P_C\approx P_{LR}^0=P_C^0 =1/2$ when $s\gg 1$, as in  Fig.~\ref{Fig:Fig5_3VMswitch}(a,b).

When $\nu\gg b$, there are many switches before reaching  the  final state (see Section~\ref{sec:METgen} below). This results in the self-averaging of $\xi$, yielding 
$\xi \to \langle \xi\rangle=0$. Hence, when $\nu\gg b$
 polarization and centrist consensus occur with probabilities:
\[P_{LR}\approx P_{LR}^{\infty}=1-\frac{1-(1-z)^2}{\sqrt{1+(1-z)^2}}  \quad \text{ and } \quad P_{C}\approx P_{C}^{\infty}=z,\] where we have used Equations 
\eqref{PLRinf} and \eqref{PCLinf} with \eqref{Pstat_nosa}. We can determine when these probabilities increase with $\nu$ by solving \eqref{zLRzC}. When $s\gg 1$, we find $z_{LR}=(4-\sqrt{18-2\sqrt{22}})/4\approx 0.3621$
and $z_C=1/2$. When $s\gg 1$
 $P_{LR}$ and $P_C$ hence increase with $\nu$ when $z<z_{LR}$
 and $z>1/2$, respectively; see Fig.~\ref{Fig:Fig5_3VMswitch}. 
 In the case $z<1/2$, $P_C$ is a decreasing function of  $\nu$ 
 and $C$ is generally the minority opinion 
  ($P_C<1/2$); see the inset in Fig.~\ref{Fig:Fig5_3VMswitch}(b).  
 As shown in Fig.~\ref{Fig:Fig5_3VMswitch}(a),
 when $z=1/2$ then  $P_C^{\infty}= P_C^{0}=1/2$, and we find that the probability of centrist consensus  remains essentially constant: $P_C(\nu)\approx 1/2$.
 When $s\gg 1,~ \delta=0$ and $z=2x=2y=1/2$, the final densities of voters are $c=P_C\approx 1/2$
 and $\ell=r= (1-P_C)/2\approx 1/4$; see the inset of Fig.~\ref{Fig:Fig5_3VMswitch}(a).


\section{Mean exit time}
\label{sec:MET}
The mean exit time (MET), here denoted by $T(\nu)$, is the average time  to reach one of three absorbing/consensus states (all-$L$, all-$R$, all-$C$) or the polarization line (pol-$LR$).
The MET hence complements the information provided by the polarization and  consensus probabilities, and is therefore of great interest.
Here, we study how the MET varies with $\nu$ for different given values of $\delta$ and $z$.

\subsection{MET in the regimes $\nu \to 0$ and $\nu \to \infty$}
\label{sec:METlimit}
The MET is obtained analytically in the regimes $\nu \to 0$ and $\nu \to \infty$
in terms of ${\cal T}(s,z)$, its counterpart in the absence of external influences, as discussed in Appendix \ref{AppB2}.

\subsubsection{MET in the regime $\nu\to 0$}
\label{sec:METzero}
When $\nu\to 0$, there are no  switches before reaching the 
final state. The MET, $T^0$, can thus be obtained by averaging  ${\cal T}(s,z)$ over the stationary distribution of $\xi$.
Since $\xi=\pm 1$ with probability $(1\pm \delta)/2$, in this regime the MET reads:
\begin{equation}
\label{Tzero}
T^0(s,\delta, z)=
\left(\frac{1+\delta}{2}\right){\cal T}(s,z)+\left(\frac{1-\delta}{2}\right){\cal T}(-s,z).
\end{equation}
$T^0(s,\delta, z)$ and its scaling are thus readily obtained from Equations \eqref{MET} and \eqref{Tq}. From the symmetry ${\cal T}(-s,z)= {\cal T}(s,1-z)$,   we have  $T^0(s,\delta,1/2)=T^0(s,-\delta,1/2)={\cal T}(s,1/2)$ for $z=1/2$~\cite{foot1}.  From Fig.~\ref{Fig:Fig6_3VMswitch},  we find that the predictions of Eq.~\eqref{Tzero} are
in good 
 agreement with simulation results when $\nu \ll 1$ 
in all scenarios $(\delta>0$, $\delta<0$ and $\delta=0$).
\subsubsection{MET in the regime $\nu\to \infty$}
\label{sec:METinf}
\begin{figure}[h!]
\begin{center}
\includegraphics[width=3.5in,angle=90]{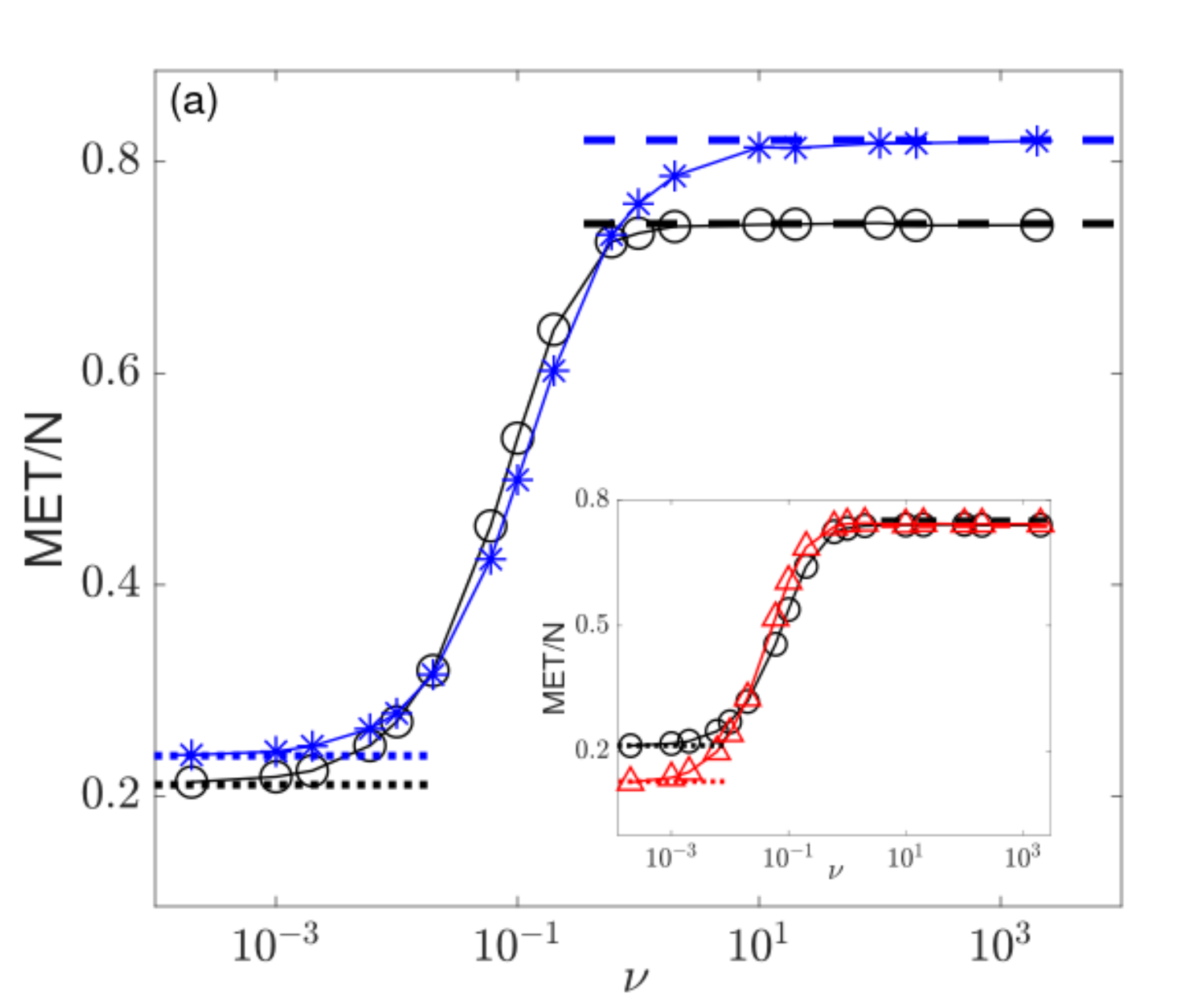}
\includegraphics[width=3.85in,angle=90]{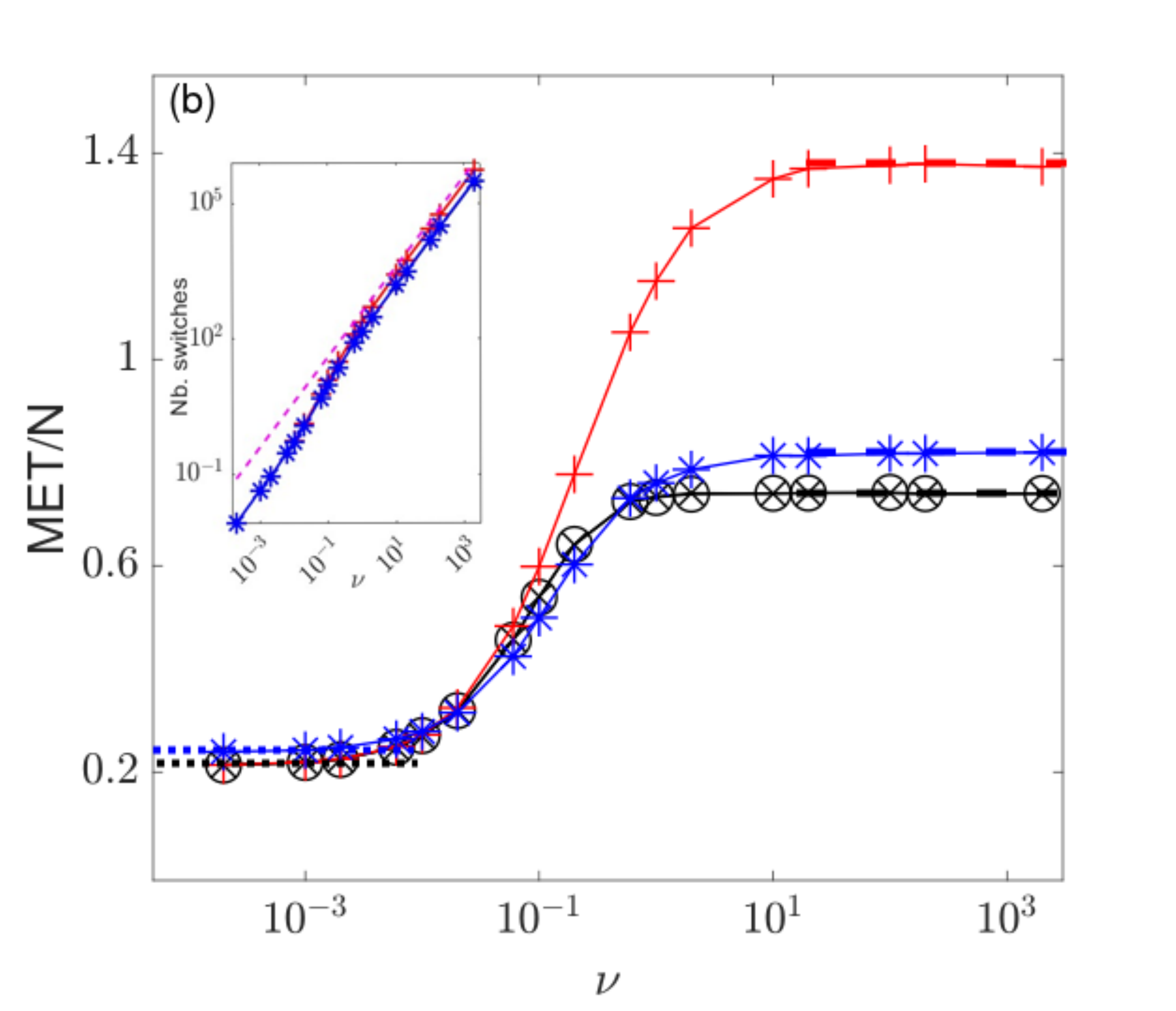}
\caption{ Scaled mean exit time
(MET) vs. $\nu$: $T(\nu)/N$ for different  $N,\delta$ and  $z$. Symbols are from simulations (averaged over $10^5$ samples).
Dotted and dashed lines
show $T^0/N$ and $T^\infty/N$ from Equations \eqref{Tzero} and \eqref{Tinf}, respectively.
({\bf a}) $(N,b,s,\delta)=(200,0.1,20,0.2)$, with  $z=0.5$ (black circles) and $z=0.88$ (blue stars).
Inset:
$T(\nu)/N$ for   $(N,b,s,\delta,z)=(500,0.08,40,0.1,0.5)$ (red triangles) and $(N,b,s,\delta,z)=(200,0.1,20,0.2,0.5)$ (black circles). MET scales as
$T(\nu)/N \sim (\ln{N})/(Nb)$ when $\nu\ll b$ and $T(\nu)/N ={\cal O}(1)$ when $\nu\gg b$; see text.
({\bf b}) $N=200,~b=0.1,~s=20$, with
$(\delta,z)=(0.2,0.5)$ (black circles), 
$(\delta,z)=(-0.2,0.5)$ (black crosses),
$(\delta,z)=(0.2,0.88)$ (blue stars), and
$(\delta,z)=(0,0.5)$ (red pluses).
Inset: average number of switches before reaching the final state vs. $\nu$ for the same parameters.
Dashed line is an eyeguide of the slope $2N\nu$. See text for details. 
}
  \label{Fig:Fig6_3VMswitch}
\end{center}
\end{figure}

When $\nu\to \infty$, many external influences switches occur  before the population reaches its final state. This leads to the self-averaging of  $\xi$, 
which can therefore be  replaced by its average: $\xi\to \langle\xi\rangle=\delta$. Hence, under high switching rate, 
the influences bias is  rescaled according to $b\to b\delta$, yielding  $s \to s\delta$, at fixed $N$. When $\nu\to \infty$ the MET, $T^\infty$, 
therefore satisfies \eqref{MET}
with $s$ substituted by $s\delta$, and thus
\begin{equation}
\label{Tinf}
T^\infty(s,\delta,z)={\cal T}(s\delta,z).
\end{equation}
In this regime, the MET  follows readily  from the solution of \eqref{MET}, and its scaling is obtained from \eqref{Tq}. Using the symmetry of ${\cal T}$,
we have  $T^\infty(s,\delta,z)=T^\infty(s,-\delta,1-z)$, which 
boils down to 
$T^\infty(s,\delta,1/2)=T^\infty(s,-\delta,1/2)$ when $z=1/2$. 
Furthermore, when $\delta \to 0$,  we obtain  $T^\infty(s,0,z)={\cal T}(0,z)=-2N \left[ z \; \ln z + (1-z) \; \ln(1-z) \right]$, which is  independent of the  influences bias $b$. 
In Fig.~\ref{Fig:Fig6_3VMswitch},  the predictions of Eq.~\eqref{Tinf}
are in close to perfect agreement with simulation results for $\nu \gg 1$
in all scenarios ($\delta>0$, $\delta<0$ and $\delta=0$).

\subsection{Mean exit time as function of the switching rate}
\label{sec:METgen}
We now consider the dependence of the MET under arbitrary $\nu$.  

When the switching rate is low, $\nu\ll b$, it is unlikely that there are any switches before reaching the final states, and thus $T(\nu)\approx 
T^0(s,\delta,z)$; see Eq.~\eqref{Tzero}. 
When $b\ll 1$,  $1\ll s\ll N$, and $z$ is not too close to $0$ and $1$, then
 $T(\nu)\sim (\ln{N})/b$ (see Eq.~\eqref{Tq}),  and 
the average number of switches in this regime prior to reaching the final state
is 
$\nu T(\nu)\sim (\ln{N})\nu/b$; see Fig.~\ref{Fig:Fig6_3VMswitch}(a,b). Hence, when $\nu\sim b$
there are on average ${\cal O}(\ln{N})$ switches before reaching the final state.

 Under high $\nu$, the system experiences a large number of switches causing  the self-average of $\xi$  (with $\xi(t)\to \delta$).
In this regime,    $T(\nu)\approx T^{\infty}(s,\delta,z)$; see Eq.~\eqref{Tinf}, with $T(\nu)\sim N$. In this case, the average number of switches before reaching the final state is $\nu T(\nu)\sim \nu N$. 

We can therefore revisit and characterise the three switching regimes:\\
(i) when $\nu\ll b$, the system is in the regime of  ``low switching rate'' and the final state can possibly be reached without experiencing any switches;\\
(ii) when $\nu \sim b$, the population is in a regime of ``intermediate switching rate'', where there are typically  ${\cal O}(\ln{N})$ switches before reaching the final state;\\ 
(iii) when $\nu\gg b$,
the system is in  the regime of  ``high switching rate'', where 
the external noise self averages, and the number of switches 
before reaching the final state is  ${\cal O}(\nu N)$ and hence grows linearly with $\nu N$, where $\nu N\gg Nb=s\gg 1$; see inset in Fig.~\ref{Fig:Fig6_3VMswitch} (b).

When the switching rate is raised across the above regimes, from $\nu \approx 0$ to $\nu \gg 1$, the MET changes its  scaling behaviour. In regime (i),  the MET scales as $T\sim (\ln{N})/b$ and therefore $T/N\ll 1$ when  $s\gg 1$, while in regime (iii)  
the MET scales as $T\sim N$, and therefore  $T/N=f(z)$ is a scaling function of the initial fraction $z$. 
In the intermediate regime (ii), when $\nu\sim b$, the MET increases steeply with $\nu$ and interpolates between $T(\nu)/N\ll 1$ and $T(\nu)/N\approx f(z)$ as $\nu$ sweeps from regimes (ii) into (iii). This picture is confirmed by the results shown in Fig.~\ref{Fig:Fig6_3VMswitch}(a,b).

Fig.~\ref{Fig:Fig6_3VMswitch}(a,b), illustrates that the initial fraction $z$ has only a marginal effect on the MET in the limiting regimes $\nu\ll b$ and $\nu\gg b$, which is well captured by 
\eqref{Tzero} and \eqref{Tinf}. Fig.~\ref{Fig:Fig6_3VMswitch}(b) shows that, as predicted by \eqref{Tinf}, the MET at fixed $N,b,z$ is maximum when $\delta=0$, with $T/N\approx T^{\infty}/N=-2 \left[ z \; \ln z + (1-z) \; \ln(1-z) \right]$; see Appendix~\ref{AppB2}. 
 The inset in Fig.~\ref{Fig:Fig6_3VMswitch}(a) illustrates that the MET scales linearly with  $N$
in the regime (iii), while $N$ has a marginal effect on $T(\nu)$ in the regimes (i) and (ii): When $\nu\gg b$, we  find
 $T(\nu)/N= {\cal O}(1)$ as in Ref.~\cite{MM2011}.
  When $\nu\ll b$, we find
 that $T(\nu)/N$  decreases with $N$ in agreement with  $T(\nu)/N\sim (\ln{N})/(Nb)$. The inset of Fig.~\ref{Fig:Fig6_3VMswitch}(b) confirms that when $\nu\gg b$, the average number of switches exhibits the same linear scaling with $\nu N$ for different parameter sets.

Here, time-varying  external influences are therefore responsible for a 
drastic change in  the MET scaling: the MET scales
as $(\ln{N})/b$ under low switching rate, while it grows and scales linearly  
with $N$ under high switching rate. When $s=Nb\gg 1$, reaching
 the final state thus takes much longer under $\nu\gg 1$ than  $\nu\ll 1$.

\section{Conclusion}
Motivated by the evolution of opinions in a volatile social environment shaped by time-fluctuating external influences, arising, e.g., from  news or social media,  we have introduced a constrained three-state voter model subject to randomly binary  time-fluctuating (switching) external influences. The voters of this population can hold either incompatible leftist or rightist opinions, or behave like centrists. The changing external influences (social environment) is modelled by a dichotomous
random process that favours in turn the rise of polarization or the spread of centrism.
The fate of this population is either to reach a consensus with leftists, rightists, or centrists, or to achieve polarization, which consists of a frozen state comprised of only non-interacting leftists and rightists. 
 By combining analytical and computational means, we have investigated the effect of the time-fluctuating external influences on the population's final state under various scenarios. In particular, we have studied how the rate of switching, as well as the switching asymmetry and initial population composition, affect the fate of this population.

 Focusing on the interesting case of a small influences bias affecting a finite, yet large, number of voters, we 
 have  shown that the consensus and polarization probabilities can vary greatly with the rate of change of the external influences: these probabilities  can either increase or decrease  with the rate of external variations, and can also exhibit extrema. Remarkably, 
  when there is a large initial majority of voters holding the opinions opposed by the external influences,  the majority can resist the influences: the opinions
  supported by the majority of agents and opposed by the influences are the most likely to prevail over a range of parameters characterising the external variations. When this happens, the population settles in a final state in which a majority of voters holds the opinions opposed by the external influences; see Fig.~\ref{Fig:Fig3_3VMswitch}(b) and Fig.~\ref{Fig:Fig4_3VMswitch}(b). The study has also shown that the time-switching  influences are  responsible for a 
drastic change in  the scaling of the mean time to reach final state:
the mean exit time is generally much bigger under high switching rate, when it scales linearly with population size, than under low switching rate.

The goal of this paper is to study how time-varying external influences may
affect the social dynamics of an idealized population. It can be anticipated that the model analysed here is too simple to realistically capture the  various complex effects of social and news media, and other time-varying external stimuli, on opinion dynamics.
It is however natural to ask how this model can be generalized to become more realistic, and which kind of practical information it could then possibly provide. These  points are addressed in Appendix \ref{AppC}, where a potential application is briefly discussed. 
In fact, while the direct applications of the current model are admittedly limited,  it is expected to still be useful since it sheds light on   nontrivial effects that exogenous time-fluctuating  influences can have on opinion dynamics. Hence, this work can be envisaged as a step towards  more realistic modelling approaches to this  challenging interdisciplinary problem. 

\vspace{3mm}
  
{\bf Data Access Statement}:  data used to generate Figures \ref{Fig:Fig3_3VMswitch} to \ref{Fig:Fig6_3VMswitch} are electronically available at the
Research Data Leeds Repository~\cite{data}.

\appendix
\section[\appendixname~\thesection]{Master equation \& mean-field limit when $N\to \infty$}
\label{AppA}

In this Appendix, we discuss the master equation (ME) governing the model's dynamics, 
and then its description in the mean-field limit when $N\to \infty$. 
\subsection{Master equation}
\label{AppA1}
The 3CVM switching dynamics in a finite population is a Markov chain 
defined by the 
transition rates:
\begin{equation}
 \label{transrate}
 \hspace{-5mm}
 W_L^{\pm}(N_L,N_R,\xi)=\frac{(1\pm b\xi)}{2}~\frac{N_L (N-N_R-N_L)}{N(N-1)}, \;\;\;
 W_R^{\pm}(N_L,N_R,\xi)=\frac{(1\pm b\xi)}{2}~\frac{N_R (N-N_R-N_L)}{N(N-1)},
\end{equation} 
such that $N_L \stackrel{W_L^{\pm}}{\longrightarrow}N_L\pm 1$ and $N_R \stackrel{W_R^{\pm}}{\longrightarrow}N_R\pm 1$. The  
associated  ME
gives the probability $P(N_L,N_R,\xi,t)$ to have $N_L$, $N_R$ and $N_C=N-N_L-N_R$
voters in the population
in
the external state $\xi=\pm1$ at time $t$~\cite{Gardiner}, and here reads:
\begin{subequations}
\label{eq:ME}
\begin{align}
\hspace{-5mm}
\frac{\partial P(N_L,N_R,\xi=1,t)}{\partial t} &=  \left( \mathbb{E}_L^--1\right)\left[W^+_L P(N_L,N_R,1,t)\right]+\left( \mathbb{E}_R^--1\right)\left[W^+_R P(N_L,N_R,1,t)\right] \nonumber \\
&+
\left( \mathbb{E}_L^+-1\right)\left[W^-_L P(N_L,N_R,1,t)\right]  +\left( \mathbb{E}_R^+-1\right)\left[W^-_R P(N_L,N_R,1,t)\right] \nonumber \\
&+ \nu\left[(1+\delta) P(N_L,N_R,-1,t)-(1-\delta) P(N_L,N_R,1,t)\right],
\\
\frac{\partial P(N_L,N_R,\xi=-1,t)}{\partial t} &=  \left( \mathbb{E}_L^--1\right)\left[W^+_L P(N_L,N_R,-1,t)\right]+\left( \mathbb{E}_R^--1\right)\left[W^+_R P(N_L,N_R,-1,t)\right] \nonumber \\
&+
\left( \mathbb{E}_L^+-1\right)\left[W^-_L P(N_L,N_R,-1,t)\right]  +\left( \mathbb{E}_R^+-1\right)\left[W^-_R P(N_L,N_R,-1,t)\right] \nonumber \\
&+ \nu\left[(1-\delta) P(N_L,N_R,1,t)-(1+\delta) P(N_L,N_R,-1,t)\right] ,
\end{align}
\end{subequations}
where $\mathbb{E}^{\pm}_{L}$ and  $\mathbb{E}^{\pm}_{R}$ are shift operators such that
$\mathbb{E}^{\pm}_{L/R} f(N_{L/R},N_{R/L},t) =f(N_{L/R}\pm 1,N_{R/L},t)$, and $P(N_L,N_R,\xi,t)=0$ whenever $N_L$ or $N_R$
are outside $[0,N]$.  $P(N_L,N_R,1,t)$ and $P(N_L,N_R,-1,t)$
are coupled, and 
the last  lines on the right-hand-side of (\ref{eq:ME}a) and (\ref{eq:ME}b)
account for the random external switching.
We notice that $W^{\pm}_{L/R}=0$ when $N_L=0, N_R=0$ and $N_L+N_R=N$, corresponding to the 
three absorbing states all-$R$ ($N_L=N_C=0$),  all-$L$ ($N_R=N_C=0$),  all-$C$ ($N_L=N_R=0$, $N_C>0$),
and to the polarization state pol-$LR$ ($N_L+N_R=N$, $N_C=0$); see Fig.~\ref{Fig:Fig2_3VMswitch}.
 The multivariate
ME \eqref{eq:ME} can be simulated exactly by standard methods, such as the Gillespie algorithm~\cite{Gillespie76}.
 
 \subsection{Mean-field limit when $N\to \infty$}
\label{AppA2}
 In the mean-field limit where $N\to \infty$ and demographic
 can be ignored, the equation of motion of the densities are~\cite{Gardiner,MM2011}:  
 \begin{equation}
 \label{SDE}
\frac{d}{dt}\ell=W_L^+-W_L^-=b\xi~\ell(1-\ell-r),\quad
\frac{d}{dt}r=W_R^+-W_R^-=b\xi~r(1-\ell-r),
\end{equation}
where  $c\equiv N_C/N=1-\ell-r$ is used, and the time dependence is omitted. Here,   the mean-field limit  means that the dynamics is aptly described by Eqs.~\eqref{SDE} 
where $\xi(t)\in\{-1,1\}$ is a randomly switching 
multiplicative noise; see Eq.~\eqref{switch}. 
Eqs.~\eqref{SDE}  are thus
two coupled {\it stochastic} differential equations that have  three absorbing steady states: 
$(\ell,r,c)=
\{(1,0,0),~(0,1,0),~(0,0,1)\}$ associated
respectively with the consensus states \text{all-}$L$, \text{all-}$R$ and \text{all-}$C$. 
Eqs. \eqref{SDE} also admit the line of steady states 
$(\ell,1-\ell,0)$, with $0<\ell<1$, associated with state of polarization \text{pol}-$LR$; see Fig.~\ref{Fig:Fig2_3VMswitch}. Eqs.~\eqref{SDE} conserve the ratio $\ell/r=x/y$, which implies 
that the final densities satisfy 
$\ell(\nu)=r(\nu)$ and $c(\nu)=1-2\ell(\nu)$ when $x=y$.

It is useful to relate this result with the final densities 
 in a \emph{finite} population, in the case where $x=y$. When $N<\infty$, the final densities
of $R$ and $L$ voters are $\ell(\nu)=r(\nu)=P_L(\nu)+P_{LR}(\nu)/2=P_R(\nu)+P_{LR}(\nu)/2$
(as in the absence of external influences~\cite{VR2004,MM2011}). The first term comes from the probability of ending in $L$ or $R$ consensus  (with $P_L=P_R$). The second term arises from the fact that, when $x=y$, the polarization state consists of half $L$ and $R$ voters.
Since $P_L+P_R+P_{LR}=1-P_C$ and $P_L=P_R$, we find 
$\ell(\nu)=r(\nu)=[1-P_C(\nu)]/2$, and the
final density of $C$ voters is:
$c(\nu)=1-2\ell(\nu)=P_C(\nu)$. 
When $N\to \infty$, Eqs.~\eqref{SDE} hold and predict that
the densities approach polarization when $\xi=1$ and centrist consensus when $\xi=-1$ on a timescale $t\sim 1/b$. Hence, when $N\to \infty$, the probability of $L$ and $R$ consensus vanish, $P_L=P_R \to 0$, and the final densities thus satisfy 
$\ell(\nu)=r(\nu)\to P_{LR}(\nu)/2$ and $c(\nu)=1-2\ell(\nu)\to 1-P_{LR}(\nu)$.

\section{Polarization, consensus probabilities, and mean exit time in the absence of time-varying influences}
\label{AppB}
In this Appendix, we reproduce
 the results obtained in Refs.~\cite{VR2004,MM2011} for the polarization and consensus probabilities in the absence of external influences, and  used in Secs.~\ref{sec:Prob} and~\ref{sec:METlimit}.
   
   \subsection{Polarization and consensus probabilities in the absence of time-varying influences}
\label{AppB1} 
    
In the realm of the diffusion theory~\cite{Gardiner,Kimura,Ewens}, when $N\gg 1$, 
$s\equiv Nb\neq 0$ and $x=y=(1-z)/2$,
 the polarization probability, here, denoted by ${\cal P}_{LR}$
 is
 ~\cite{MM2011}:
\begin{align}
\label{Pab-stat}
{\cal P}_{LR}(s,z)=\frac{e^{sz}\sqrt{1-z}}{2}\, &\sum_{k=0}^{\infty}\frac{(-1)^k(4k+3)}{(2k+1)(k+1)}~\frac{(2k+1)!!}{(2k)!!}
\frac{I_{2k+3/2}(s(1-z))}{I_{k+3/2}(s)}, 
\end{align}
where $I_k(\cdot)$ denotes the modified Bessel function of first kind and order $k$, and we  used 
 the property 
 $P_{2k+1}^{1}(0)=(-1)^k{(2k+1)!!}/{(2k)!!}$
 of the associated Legendre polynomials, $P_{l}^m(x)$~\cite{foot3}.    
 When 
 $s<0$ and $|s|(1-z)\gg 1$, Eq.~\eqref{Pab-stat} can be approximated by ${\cal P}_{LR}\approx (e^{2|s|(1-z)}-1)/(e^{2|s|}-1)$;  
 see Ref.~
\cite{MM2011}. This simplified expression is particularly useful to approximate $P_{LR}^{\infty}$ when $\delta<0$; 
see Eq.~\eqref{PLRinf}.
 
 In the realm of the diffusion theory,  when $N\gg 1$, 
$s\equiv Nb\neq 0$, the $C$-consensus
 probability, here denoted by ${\cal P}_{C}$, is
 ~\cite{MM2011}:
\begin{align}
\label{Pc-stat}
{\cal P}_{C}(s,z)=\frac{e^{-2s(1-z)}-e^{-2s}}{1-e^{-2s}}.
\end{align}

When $z\neq 0,1$, one has $\lim_{s\to \infty}{\cal P}_{LR}(s,z)= 1$ and $\lim_{s\to -\infty}{\cal P}_{C}(s,z)= 1$. In the considered examples, when  $s\gg 1$ and $z$ is not too close to $0$ or $1$, polarization and centrist consensus are almost certain, i.e. ${\cal P}_{LR}(s,z)\approx 1$
if $s>0$ and ${\cal P}_{C}(s,z)\approx 1$ when $s<0$~\cite{MM2011}.

When  $b=s=0$,
and $x=y=(1-z)/2$, the probabilities ${\cal P}_{LR}$ and ${\cal P}_{C}$ become~\cite{VR2004}:
\begin{equation}
\label{Pstat_nosa}
{\cal P}_{LR}(0,z)=1-\frac{1-(1-z)^2}{\sqrt{1+(1-z)^2}}, \quad
{\cal P}_{C}(0,z)=z.
\end{equation}
When $x=y=(1-z)/2$, 
the probability ${\cal P}_{R}={\cal P}_{L}$ to end up in an $L$ or $R$ consensus  is thus
\begin{align}
 \label{PA_stat}
 {\cal P}_{R}(s,z)&={\cal P}_{L}(s,z)=\frac{1-{\cal P}_{LR}(s,z)-{\cal P}_{C}(s,z)}{2},
\end{align}
while there is the same average final fraction 
$\ell$ and $r$ of voters of type $L$ and $R$, given by 
\begin{align}
 \label{lr_stat}
r&=\ell=\frac{{\cal P}_{LR}(s,z)}{2}+{\cal P}_{R}(s,z)=\frac{1-{\cal P}_{C}(s,z)}{2}.
\end{align}
Expressions \eqref{PA_stat} and \eqref{lr_stat} hold both when $s\neq 0$ with \eqref{Pab-stat} and \eqref{Pc-stat}, and when $s=0$ with \eqref{Pstat_nosa}.

   
   \subsection{Mean exit time in the absence of time-varying influences}
\label{AppB2}    
In Refs.~\cite{VR2004,MM2011}, the (unconditional)
mean exit time 
of the 3CVM in the absence of time-varying influences, here denoted by ${\cal T}$,  was shown to satisfy: 
\begin{equation}
\label{MET}
\frac{z(1-z)}{2N}\left[-2s\frac{d{\cal T}(s,z)}{dz} + \frac{d^2 {\cal T}(s,z)}{dz^2} \right]=-1,
\end{equation}
with ${\cal T}(s,0)={\cal T}(s,1)=0$
~\cite{foot4}. 
The symmetry of \eqref{MET} under
$(s,z)\to (-s,1-z)$ implies ${\cal T}(-s,z)={\cal T}(s,1-z)$~\cite{MM2011}.
When $s=Nb\neq 0$ and $|b|\ll 1$, with  $z$  not too close to $0$ and $1$, the MET scaling is~\cite{Blythe07,Ewens,MM2011}:
\begin{eqnarray} 
\label{Tq}
{\cal T}(s,z)&\sim&
\begin{cases}
 \frac{N\ln{N}}{|s|} & \text{when }~ 1 \ll |s|\ll N \\
 N &  \text{when $|b|\ll 1$ and $|s|= {\cal O}(1)$},
\end{cases}
\end{eqnarray}
yielding ${\cal T}(s,z)={\cal O}(\ln{N}/b)$ when  $|b|\ll 1$ and $1 \ll |s|\ll N$.  When $b=s=0$, the MET ${\cal T}(0,z)$ takes the simple closed
form ${\cal T}(0,z)=-2N \left[ z \; \ln z + (1-z) \; \ln(1-z) \right]$~\cite{VR2004},
%
and in this case, it scales linearly with the population size: ${\cal T}(0,z)\sim N$.
\section{Possible generalizations and applications of the model}

\label{AppC}

This Appendix is dedicated to a brief discussion of some possible generalizations and applications of the 3CVM with switching dynamics.

It is often hard  to map 
complex real social systems onto idealized
theoretical models: these commonly suffer from a number of limitations that
make their use for the practical characterization of social behaviour difficult. The 3CVM  with switching dynamics is  no exception, and among its limitations, we can list the following: (i) it is doomed to end up in either consensus or $L/R$-polarization, but never admits the long-lived coexistence of the three opinions; (ii) the present model formulation assumes that all agents interact with all others on a complete graph (rather than on a complex dynamic network); (iii) as most classical voter models,  all agents are identical in the 3CVM (there are no ``zealots''); (iv) interactions are pairwise (no group pressure); (v) in many applications, there are more than three possible opinions/parties. 

Since the 3CVM can be generalized to overcome the limitations (i)-(v) at the expense of its mathematical tractability, it is useful to discuss a possible application. The main challenge for this is to find data against which to calibrate the parameters $\nu$ and $\delta$ characterising the external time-varying influences. In this context, the formulation of the 3CVM suggests to test its use to describe
the distribution of opinions in the
readership of a newspaper such as the \emph{Guardian} in the UK whose political backing of the Labour, Liberal (Lib Dem) and Conservative parties  has changed on various occasions in the last 78 years.

We have used  the dataset~\cite{dataGuardian} referring to 
the 18 general elections held in the UK between 1945-2010~to try and estimate the parameters  $\nu$ and $\delta$ for a population corresponding to a random sample of the readership of the \emph{Guardian}  (whose circulation between 1945-2021 has varied between $10^5$ and $3 \times 10^5$). If we use the average time between each general election
as unit of time, and notice that the \emph{Guardian}'s political orientation changed  8 times between 1945-2010, in 1950, 1951, 1955, 1959, 1974, 1979, 2005, 2010. The average switching rate can thus be estimated as $\nu\approx 8/18$. By treating the backing of the Labour or Conservative party as being in the influences state $\xi=1$, and representing the backing of the Lib Dem party by $\xi=-1$, we estimate that on average $\langle \xi \rangle=\delta \approx 6/18$.
Note that here the change from backing jointly two parties (e.g., Labour and Lib Dem) 
to supporting only one of those parties
(e.g., Liberal Party) is considered as a ``switch''~\cite{foot5}.
 The parameter $b$ could in principle be estimated from the approach to the ``final state'' occurring  on a timescale $\sim 1/b$; see \eqref{SDE}. In this context, it is a difficult task to assess on what timescale  consensus or polarization may occur, if ever. Here,  for the sake of argument,  we set $b=0.01$. With $(b,\nu,\delta)=(0.01,0.44,0.33)$
and assuming an initial population consisting of $60\%$ and $10\%$ of labour and conservative supporters, respectively, and $30\%$ of Lib Dem backers, the model predicts a final state consisting of
$70\%$ and  $12\%$ of Labour and Conservative supporters, respectively, and $18\%$ of  Lib Dem backers in a random sample  of size $N=200$. For a larger  sample of size $N=1000$ and the same above parameters, the model predicts a final
state comprised of
$85\%$ and  $14\%$ of Labour and Conservative supporters, and only 
a small fraction of $1\%$  backing the Lib Dem party. These figures, suggesting the slow evolution towards  quasi polarization of the \emph{Guardian}'s readership, with a raise of support for the Labour party,  as $N$ increases, do not seem absurd but are  not realistic as they  underestimate the Lib Dem vote. Moreover, the 3CVM ignores entirely the existence of a small but non-negligible fraction of
voters backing other parties (such as the Green party). A more realistic model  would take into account more than three parties, and
mechanisms ensuring the maintenance of long-lived coexistence of all opinions.


\begin{thebibliography}{99}

%
\bibitem{Granovetter}
Granovetter, M.
Threshold Models of Collective Behavior,
 \href{https://www.journals.uchicago.edu/doi/abs/10.1086/226707}{{\em Am.\ J. Sociol.} {\bf 1978}, {\em 83},  1420--1443}.
%
\bibitem{Schelling} 
Schelling, T.~C. \href{https://www.scribd.com/document/361040557/Schelling-T-Micromotives-and-macrobehavior-1978-pdf}{{\em Micromotives \& Macrobehaviour}}; W. W. Norton \& Company, Inc: New York, NY, USA, 1978. 
%
\bibitem{Galam1}
Galam, S.; Gefen, Y; Shapir, Y. 
Sociophysics: A new approach of sociological collective behaviour: I. Mean-behaviour description of a strike.
\href{https://doi.org/10.1080/0022250X.1982.9989929}{Math. {\em J. of Sociology}
{\bf 1982}, {\em 9}, 1--13}.
%
\bibitem{Galam2}
Galam, S. Majority rule, hierarchical structures, and democratic totalitarianism: A statistical approach.  
\href{https://doi.org/10.1016/0022-2496(86)90019-2}{{\em J. Math. Psychology} {\bf 1986}, {\em 30}, 426--434}.
%
\bibitem{Galam3}
Galam, S. Social paradoxes of majority rule voting and renormalization group. 
\href{https://doi.org/10.1007/BF01027314}{{\em J.~Stat.~Phys.} {\bf 1990}, {\em 61}, 943--951}.
%
\bibitem{Galam4}
Galam, S.; Moscovici, S. Towards a theory of collective phenomena: Consensus and attitude changes in groups.  
\href{https://doi.org/10.1002/ejsp.2420210105}{{\em Eur. J. Soc. Psychol.} {\bf 1991}, {\em 21}, 49--74}.
%
\bibitem{Castellano-rev}
Castellano, C.; Fortunato, S.; Loreto, V.
Statistical physics of social dynamics.
\href{https://doi.org/10.1103/RevModPhys.81.591}{{\em Rev. Mod. Phys.} {\bf 2009}, {\em 81}, 591--646}.
%
\bibitem{Galam-book}
Galam, S.
{\em Sociophysics. A Physicist's Modeling of Psycho-political Phenomena}; Springer Science+Business Media, LLC: New York, NY, USA, 2012.
%
\bibitem{Sen-book}
Sen, P; Chakrabarti, B.~K.
{\em Sociophysics: An Introduction}; Oxford University Press: New York, NY, USA, 2014.
%
%
\bibitem{Perc-2017}
Perc, M.; Jordan, J.~J.;  Rand, D.~G.; Wang, Z.; Boccaletti, S.; Szolnoki, A.
Statistical physics of human cooperation.
\href{https://doi.org/10.1016/j.physrep.2017.05.004}{{\em Physics Reports}
{\bf 2017}, {\em 687}, 1--51}.
%
\bibitem{Schweitzer2018}
Schweitzer, F. Sociophysics.
\href{https://doi.org/10.1063/PT.3.3845}{{\em Physics Today} {\bf 2018}, {\em 71}, 40--46}.
%
\bibitem{SW-2019}
Jedrzejewski, A.; Sznajd-Weron, K. Statistical physics of opinion formation: is it a SPOOF?  \href{https://doi.org/10.1016/j.crhy.2019.05.002}{{\em C. R. Physique}
{\bf 2019}, {\em 20}, 244--261}.
%
\bibitem{Redner-2019}
Redner, S. 
Reality-inspired voter models: A mini-review. 
 \href{https://doi.org/10.1016/j.crhy.2019.05.004}{{\em C. R. Physique}
{\bf 2019}, {\em 20}, 275--292}.
%
%
\bibitem{Szolnoki-2022}
Li, Z.; Chen, X.;  Yang, H.-X.;  Szolnoki, A.
Game-theoretical approach for opinion dynamics on social networks.
\href{https://doi.org/10.1063/5.0084178}{{\em Chaos}
{\bf 2022}, {\em 32}, 73117:1-13}.
%
\bibitem{Liggett}
Liggett, T.~M. \href{https://doi.org/10.1007/b138374}{{\em Interacting Particle Systems}}; Springer: Berlin/Heidelberg, 2005. 
%
\bibitem{Glauber}
Glauber, R.~J. Time‐Dependent Statistics of the Ising Model.  
\href{https://doi.org/10.1063/1.1703954}{{\em J. Math. Phys.} {\bf 1963}, {\em 4}, 294--307}.
%
\bibitem{Asch}
Asch, S.~E. Opinions and Social Pressure. 
\href{https://www.scientificamerican.com/article/opinions-and-social-pressure/}{{\em Scientific American} {\bf 1955}, {\em 193}, 31--35}. 
\bibitem{Milgram}
Milgram, S.; Bickman, L;  Berkowitz, L. Note on the drawing power of crowds of different size. \href{https://doi.org/10.1037/h0028070}{{\em J. of Personality and Soc. Psychology} {\bf 1969}, {\em 13}, 79--82}.
%
\bibitem{Lattane}
Lattan\'e, B. The psychology of social impact. 
\href{https://doi.org/10.1037/0003-066X.36.4.343}{{\em Am. Psychologist} {\bf 1981}, {\em 36}, 343--356}.
%
\bibitem{Axelrod}
 Axelrod, R. The Dissemination of Culture: A Model with Local Convergence and Global Polarization.
 \href{https://doi.org/10.1177/0022002797041002001}{{\em J. Conflict Resolution} {\bf 1997}, {\em 41}, 203--226}.
 %
 \bibitem{Axelrod-book}
 Axelrod, R. \href{https://doi.org/10.1515/9781400822300}{{\it The complexity of cooperation:  Agent-Based Models of Competition and Collaboration}}, Princeton University
Press, Princeton, USA, 1997. 
%
\bibitem{Castellano2000}
Castellano, C.;
Marsili, M; Vespignani, A. Nonequilibrium Phase Transition in a Model for Social Influence.
\href{https://doi.org/10.1103/PhysRevLett.85.3536}{{\em Phys. Rev. Lett.} {\bf 2000}, {\em 85}, 3536--3539}.
%
\bibitem{Klemm2003}
Klemm, K.;  Eguiluz, V.~M.; Toral, R;  San Miguel, M. 
Global culture: A noise-induced transition in finite systems.
\href{https://doi.org/10.1103/PhysRevE.67.045101}{{\em Phys.~Rev.~E} {\bf 2003}, {\em 67}, 045101:1--4}.
%
\bibitem{McPherson1987}
McPherson J.~M.; Smith-Lovin, L.
Homophily in voluntary organizations: Status distance and the composition of face-to-face groups.
\href{https://doi.org/10.2307/2095356}{{\em Am. Sociol. Rev. } {\bf 1987},
{\em 52}, 370--379}.
%
\bibitem{Mobilia2003}
Mobilia, M. Does a Single Zealot Affect an Infinite Group of Voters? \href{https://doi.org/10.1103/PhysRevLett.91.028701}{{\em Phys. Rev. Lett.} {\bf 2003}, {\em 91}, 028701:1--4}.
%
\bibitem{Mobilia2005}
Mobilia, M; Georgiev, I.~T. 
Voting and catalytic processes with inhomogeneities.
\href{https://doi.org/10.1103/PhysRevE.71.046102}{{\em Phys. Rev. E} {\bf 2005},
{\em 71}, 046102:1--17}. 
%
\bibitem{Mobilia2007}
Mobilia, M.; Petersen, A.;  Redner, S. 
On the role of zealotry in the voter model.
\href{https://doi.org/10.1088/1742-5468/2007/08/P08029}{{\em J. Stat. Mech.} {\bf 2007}, {\em P08029}, 1--17}.
%
\bibitem{Mobilia2013}
Mobilia, M. Commitment versus persuasion in the three-party constrained voter model. \href{https://doi.org/10.1007/s10955-012-0656-x}{{\em J. Stat. Phys.} {\bf 2013}, {\em 151}, 69--91}.
%
\bibitem{Mobilia2015}
Mobilia, M. Nonlinear $q$-voter model with inflexible zealots. 
\href{https://doi.org/10.1103/PhysRevE.92.012803}{{\em Phys. Rev. E} {\bf 2015}, {\em 92}, 012803:1--11}.
%
\bibitem{zealotother1}
Galam, S; Jacobs, F. 
The role of inflexible minorities in the breaking of democratic opinion dynamics. \href{https://doi.org/10.1016/j.physa.2007.03.034}{{\em Physica A} {\bf 2007}, {\em 381}, 366--376}.
%
\bibitem{zealotother2}
Sznajd-Weron, K.; Tabiszewski, M.;  Timpanaro, A.~M.
Phase transition in the Sznajd model with independence.
\href{https://doi.org/10.1209/0295-5075/96/48002}{{\em EPL (Europhys. Lett.)} {\bf 2011}, {\em 96}, 48002:p1--p6}.
%
\bibitem{zealotother3}
Acemoglu, D; Como, G.; Fagnani, F.;  Ozdaglar, A.
Opinion fluctuations and disagreement in social networks.
\href{https://doi.org/10.1287/moor.1120.0570}{{\em Math. Op. Res.} 
{\bf 2013},
{\em 38}, 1--27}.
%
\bibitem{zealotother4}
Nyczka, P.; Sznajd-Weron, K. 
Anticonformity or Independence?—Insights from Statistical Physics.
\href{https://doi.org/10.1007/s10955-013-0701-4}{{\em J. Stat Phys.} {\bf 2013}, {\em 151}, 174--202}.
%
\bibitem{zealotother5}
Masuda, N. Opinion control in complex networks
\href{https://doi.org/10.1088/1367-2630/17/3/033031}{{\em New J. Phys.} {\bf 2015}, {\em 17}, 033031:1--11}.
%
\bibitem{zealotother6}
Li, X; Mobilia, M.; Rucklidge, A.~M.; Zia, R.~K.~P. How does homophily shape the topology of a dynamic network?
\href{https://doi.org/10.1103/PhysRevE.104.044311}{{\em Phys. Rev. E} {\bf 2021}, 
{\em 104}, 044311:1--11}.
%
\bibitem{zealotother7}
Li, X; Mobilia, M.; Rucklidge, A.~M.; Zia, R.~K.~P. Effects of homophily and heterophily on preferred-degree networks: mean-field analysis and overwhelming transition
\href{https://doi.org/10.1088/1742-5468/ac410f}{{\em J. Stat. Mech.} {\bf 2022}, {\em 013402}, 1--27}.
%
\bibitem{qVM}
Castellano, C; Mu\~noz, M.~A.;  Pastor-Satorras, R.
Nonlinear $q$-voter model.
\href{https://doi.org/10.1103/PhysRevE.80.041129}{{\em Phys. Rev.
E} {\bf 2009}, {\em 80}, 041129:1--8}.
%
\bibitem{Sznajd}
Sznajd-Weron, K.;  Sznajd, J. 
Opinion evolution in closed community.
\href{https://doi.org/10.1142/S0129183100000936}{{\em Int.~J.~Mod.~Phys. C} {\bf 2000}, {\em 11}, 1157--1165}.
%
\bibitem{Slanina} 
Slanina F.; Lavicka, H. 
Analytical results for the Sznajd model of opinion formation.
\href{https://doi.org/10.1140/epjb/e2003-00278-0}{{\em Eur. Phys. J. B} {\bf 2003}
{\em 35}, 279--288}.
%
\bibitem{genSznajd}
Nyczka, P.; Sznajd-Weron, K.; Cislo, J.
Opinion dynamics as a movement in a bistable potential.
\href{https://doi.org/10.1016/j.physa.2011.07.050}{{\em Physica A} 
{\bf 2012},
{\em 391}, 317--327}.
%
\bibitem{vacillating}
 Lambiotte, R.; Redner, S. Dynamics of non-conservative voters. \href{https://doi.org/10.1209/0295-5075/82/18007}{{\em EPL (Europhys. Lett.)} {\bf 2008}, {\em 82}, 18007:p1--p5}.
%
\bibitem{exitprobq2} 
Slanina, F.; Sznajd-Weron, K.;  Przybyla, P. Some new results on one-dimensional outflow dynamics.
\href{https://doi.org/10.1209/0295-5075/82/18006}{{\em EPL (Europhys. Lett.)} {\bf 2008},
{\em 82}, 18006:p1--p6}.
%
\bibitem{Galam6}
Galam, S;  Martins, A.~C.~R.
Pitfalls driven by the sole use of local updates in dynamical systems.
\href{https://doi.org/10.1209/0295-5075/95/48005}{{\em EPL (Europhys. Lett.)} 
{\bf 2011},
{\em 95}, 48005:p1--p5}.
%
\bibitem{exitprobq}
Timpanaro, A.~M.;  Prado, C.~P.~C.
Exit probability of the one-dimensional q-voter model: Analytical results and simulations for large networks. 
\href{https://doi.org/10.1103/PhysRevE.89.052808}{{\em Phys. Rev.
E} {\bf 2014}, {\em 89}, 052808:1--8}.
%
\bibitem{Mellor2016}
Mellor, A.; Mobilia, M.;  Zia, R.~K.~P. Characterization of the nonequilibrium steady state of a heterogeneous nonlinear $q$-voter model with zealotry.  \href{https://doi.org/10.1209/0295-5075/113/48001}{{\em EPL (Europhys. Lett.)} {\bf 2016}, {\em 113}, 48001:p1--p6}.
%
\bibitem{Mellor2017}
Mellor, A.; Mobilia, M.;  Zia, R.~K.~P. Heterogeneous out-of-equilibrium nonlinear $q$-voter model
with zealotry. \href{https://doi.org/10.1103/PhysRevE.95.012104}{{\em Phys. Rev. E} {\bf 2017}
{\em 95}, 012104:1--15}.
%
%
\bibitem{BoundedCompromise1}
Deffuant, G.; Neau, D.;  Amblard, F.;  Weisbuch, G. Mixing Beliefs among Interacting Agents. 
\href{https://doi.org/10.1142/S0219525900000078}{{\em Adv. Complex Syst.} {\bf 2000}, {\em 3}, 87--98}.
%
\bibitem{BoundedCompromise2}
Hegselmann, R.; Krause, U. Opinion dynamics and bounded confidence models, analysis, and simulation.
\href{http://jasss.soc.surrey.ac.uk/5/3/2.html}{{\em J. Artif. Soc. Soc. Simul.} {\bf 2002} {\em 5}, Issue No 3}.
%
\bibitem{BoundedCompromise3}
Weisbuch, G.; Deffuant, G.; Amblard, F.;  Nadal, J.~P. Meet, discuss, and segregate! \href{https://doi.org/10.1002/cplx.10031}{{\em Complexity} {\bf 2002}, {\em 7}, 55--63}.
%
\bibitem{BoundedCompromise4}
Ben-Naim, E.; Krapivsky, P.~L.; Redner, S. Bifurcations and Patterns in Compromise Processes, 
\href{https://doi.org/10.1016/S0167-2789(03)00171-4}{{\em Physica D} {\bf 183}, 190 (2003)}.
%
\bibitem{BoundedCompromise5}
Slanina, F. Dynamical phase transitions in Hegselmann-Krause model of opinion dynamics and consensus.  \href{https://doi.org/10.1140/epjb/e2010-10568-y}{{\em Eur. Phys. J. B} {\bf 2011}, {\em 79}, 99--106}.
%
%
\bibitem{VKR2003}
Vazquez, F.;  Krapivsky, P.~L.; Redner, S.
Freezing and Slow Evolution in a Constrained Opinion Dynamics Model. \href{https://doi.org/10.1088/0305-4470/36/3/103
}{{\em J.~Phys.~A:~Math.~Gen.} {\bf 2003},
{\em 36}, L61--L68}.
%
\bibitem{VR2004}
Vazquez, F.;  Redner, S. Ultimate fate of constrained voters. \href{https://doi.org/10.1088/0305-4470/37/35/006}{{\em J.~Phys.A:~Math.~Gen.} {\bf 2004}, {\em 37}, 8479--8494}.
%
\bibitem{MM2011}
Mobilia, M. Fixation and polarization in a three-species opinion dynamics model, \href{https://doi.org/10.1209/0295-5075/95/50002}{{\em EPL (Europhys. Lett.)} {\bf 2011}, {\em 95}, 50002:p1--p6}.
%
%
\bibitem{Galam23}
Galam, S. Unanimity, Coexistence, and Rigidity: Three Sides of Polarization,
 \href{https://doi.org/10.3390/e25040622}{{\em Entropy} {\bf 2023}, {\em 25}, 622:1--15}.
%
\bibitem{Media1}
 DellaVigna, S;  Kaplan, E. The Fox News Effect: Media Bias and Voting. \href{https://doi.org/10.1162/qjec.122.3.1187}{{\em Quart. J. Econ.} {\bf 2007}, {\em 122}, 1187--1234}.
 %
\bibitem{Media2}
 Gerber, A.~S.;  Karlan, D.;  Bergan, D.
 Does the Media Matter? A Field Experiment
Measuring the Effect of Newspapers
on Voting Behavior and Political Opinions. \href{https://www.aeaweb.org/articles?id=10.1257/app.1.2.35}{{\em American Economic Journal: Applied Economics}, {\bf 2009}, {\em 1}, 35--52}.
 %
 \bibitem{Media3}
Yang, J.;  Leskovec, J. Patterns of Temporal Variation in Online Media. \href{https://doi.org/10.1145/1935826.1935863}{in {\em WSDM'11: Proceedings of 
the 
Fourth ACM 
International Conference on Web Search and Data Mining, Hong Kong, 9--12 February 2011}}; King, I., 
Nejdi, W., Li, H., Eds.; Association for Computing Machinery, New York, USA, 2011; pp. 177--186.
%
 \bibitem{Media4}
  Wettstein, M.;  Wirth, W. Media Effects: How Media Influence Voters. \href{https://doi.org/doi:10.1111/spsr.12263}{{\em Swiss Polit. Science Review} {\bf 2017},
  {\em 23}, 262--269}.
  %
 \bibitem{Media5}
 Holbrook, T.~M.   \href{https://doi.org/10.1093/acprof:oso/9780190269128.001.0001}{{\it Altered States: Changing Populations, Changing Parties, and the Transformation of the American
Political Landscape}}, Oxford University Press: New York, USA, 2016.
%
\bibitem{Media6}
Dewenter, R.; Linder, M.; Thomas, T.
 Can media drive the electorate? The impact of media coverage on voting intentions. \href{https://doi.org/10.1016/j.ejpoleco.2018.12.003}{{\em Eur. J. 
Polit. 
Econ.} {\bf 2019}, {\em 58}, 245--261}.
%
\bibitem{Media7}
 Kaviani, M.~S.; Li, L.; Maleki, H.
Media, Partisan Ideology, and Corporate Social Responsibility. \href{https://doi.org/10.2139/ssrn.3658502}{\emph{SSRN} {\bf 2021}, 3658502.}.
%
\bibitem{Bhat1}
 Bhat, D;   Redner, S. Nonuniversal opinion dynamics driven by opposing external influences. \href{https://doi.org/10.1103/PhysRevE.100.050301}{{\em Phys. Rev. E} {\bf 2019}, {\em 100}, 050301(R):1--5}. 
%
\bibitem{Bhat2}
Bhat, D;   Redner, S. Polarization and consensus by opposing external sources. \href{https://doi.org/10.1088/1742-5468/ab6094}{{\em J. Stat. Mech.} {\bf 2020}, {\em 013402}, 1--28}. 
%
\bibitem{NatureCom2019}
 Lorenz-Spreen, P.; M\o{}rch M\o{}nsted, B.;  H\"ovel, P.;  Lehmann, S. Accelerating dynamics of collective attention.
\href{https://doi.org/10.1038/s41467-019-09311-w}{{\em Nat. Commun.} {\bf 2019}, {\em 10}, 1759:1--9}.
%
\bibitem{Media23}
 Helfmann, L.;  Djurdjevac Conrad, N.;  Lorenz-Spreen, P.;   Sch\"utte, C. Modelling opinion dynamics
under the impact of influencer and media strategies. \href{https://doi.org/10.48550/arXiv.2301.13661}{{\em E-print: arXiv:2301.13661}  {\bf 2023}}.

\bibitem{Hyst2}
Hosseiny, A.; Bahrami, M.;  Palestrini, A.; Gallegati, M.
Metastable features of economic networks
and responses to exogenous shocks.
\href{https://doi.org/10.1371/journal.pone.0160363}{{\em PLoS ONE} {\bf 2016}, {\em 11}, 1--22}. 
%
\bibitem{Hyst3}
Hosseiny, A.; Absalan, M.;  Sherafati, M.; Gallegati, M.
Hysteresis of economic networks in an XY model.
\href{https://doi.org/10.1016/j.physa.2018.08.064}{{\em Physica A} {\bf 2019}, {\em 513}, 644--652}. 


%

\bibitem{Kimura}  Crow, J.~F.;  Kimura, M. \href{https://archive.org/details/an-introduction-to-population-genetics-theory-pdfdrive}{{\it An Introduction to Population Genetics Theory}}. The Blackburn Press, New Jersey, USA, 1970.
%
\bibitem{Ewens}
 Ewens, W.~J. {\it Mathematical Population Genetics. Volume 1: Theoretical Introduction}. Springer Science+Business Media New York: New York, NY, USA, 2004.
%
\bibitem{AMR2013}
Assaf, M.;  Mobilia, M.;  Roberts, E. Cooperation Dilemma in Finite Populations under Fluctuating Environments. \href{https://doi.org/10.1103/PhysRevLett.111.238101}{{\em Phys. Rev. Lett.} {\bf 2013}, {\em 111}, 238101:1--5}.
%
\bibitem{WMR2018}
 West, R.; Mobilia, M.;  Rucklidge,  A.~.M. Survival behavior in the cyclic Lotka-Volterra model with a randomly switching reaction rate,\href{https://journals.aps.org/pre/abstract/10.1103/PhysRevE.97.022406}{Phys. Rev. E. {\bf 97}, 022406  (2018)}.
%
\bibitem{HL06}
Horsthemke, W.; Lefever, R. {\it Noise-Induced Transitions}. Springer, Berlin, Germany, 2006.
%
\bibitem{Bena2006}
Bena, I. Dichotomous noise: exact results for out-of-equilibrium systems. 
\href{https://doi.org/10.1142/S0217979206034881}{{\em Int.~J.~Mod.~Phys. B} {\bf 2006}, {\em 20}, 2825--2888}.
%
\bibitem{Ridolfi11}
Ridolfi, L.; D'Odorico, P.;  Laio, F. \href{https://doi.org/10.1017/CBO9780511984730}{{\it Noise-Induced Phenomena in the environmental Sciences}}. Cambridge University Press: Cambridge, UK, 2011.
%
\bibitem{KEM1}
Wienand, K.; Frey, E.; Mobilia, M. Evolution of a Fluctuating Population in a Randomly Switching Environment. \href{https://doi.org/10.1103/PhysRevLett.119.158301}{{\em Phys.~Rev.~Lett.} {\bf 2017}, {\em 119}, 158301:1--65}.
%
\bibitem{KEM2}
Wienand, K.; Frey, E.; Mobilia. Eco-evolutionary dynamics of a population with randomly switching carrying capacity. \href{https://doi.org/10.1098/rsif.2018.0343}{{\em J. R.~Soc.~Interface} {\bf 2018}, {\em 15}, 20180343:1--12}.
%
\bibitem{TWAM}
Taitelbaum, A.; West, R.; Assaf, M.; Mobilia, M. Population Dynamics in a Changing Environment: Random versus Periodic Switching. \href{https://doi.org/10.1103/PhysRevLett.125.048105}
{{\em Phys.~Rev.~Lett.} {\bf 2020}, {\em 125}, 048105:1--6}.
%
\bibitem{TWAM2}
Taitelbaum, A.; West, R.; Mobilia, M; Assaf, M. Evolutionary Dynamics in a Varying Environment: Continuous versus Discrete Noise. \href{https://doi.org/10.1103/PhysRevResearch.5.L022004}
{{\em Phys.~Rev.~Research.} {\bf 2023}, {\em 5}, L022004:1--7}.
%
%
\bibitem{Shnerb2015}
Kalyuzhny, M.; Kadmon, R.; Shnerb, N.~M. A neutral theory with environmental stochasticity explains static and dynamic properties of ecological communities. 
\href{http://dx.doi.org/10.1111/ele.12439}
{{\em Ecology Letters} {\bf 2015}, {\em 18}, 572--580}.
%
%
\bibitem{Hufton2016}
Hufton, P.-G.; Lin, Y.-T.; Galla, T.; McKane, A.~J. Intrinsic noise in systems with switching environments. \href{https://doi.org/10.1103/PhysRevE.93.052119}{{\em Phys.~Rev.~E} {\bf 2016}, {\em 93}, 052119:1--13}.
%
\bibitem{Hidalgo2017}
Hidalgo, J.; Suweis, S.; Maritan, A. Species coexistence in a neutral dynamics with environmental noise. \href{https://doi.org/10.1016/j.jtbi.2016.11.002}{{\em J.~Theor.~Biol.}
{\bf 2017}, {\em 413}, 1--10}.
%
%
\bibitem{Gardiner}  Gardiner, C.~W. {\it Stochastic Methods: A Handbook for the Natural and Social Sciences}, Springer: Berlin/Heidelberg, Germany, 2009.
%
\bibitem{Hyst1}
Chakrabarti, B. K.; Acharyya, M.
Dynamic transitions and hysteresis.
\href{https://doi.org/10.1103/RevModPhys.71.847}{{\em Rev. Mod. Phys.} {\bf 1999}, {\em 71}, 847--859}. 
%
\bibitem{data}
Mobilia, M. {\it SimData$\_$Figs3to6} [Dataset]. University of Leeds: Leeds, UK, 2023. \href{https://doi.org/10.5518/1311}{https://doi.org/10.5518/1311}.
%
\bibitem{Gillespie76}
Gillespie, D.~T. A general method for numerically simulating the stochastic time evolution of coupled chemical reactions. 
\href{https://doi.org/10.1016/0021-9991(76)90041-3}{{\em J. Comput. Phys.} {\bf 1976} {\em 22}, 403--434}.
%
%
\bibitem{Blythe07}
Blythe, R.~A.;  McKane, A.~J. Stochastic models of evolution in genetics, ecology and linguistics. \href{https://doi.org/10.1088/1742-5468/2007/07/P07018}{{\em J. Stat. Mech.} {\bf 2007}, {\em P07018}, 1--58}.
%
\bibitem{dataGuardian}
\emph{The Guardian}. Datablog. Newspaper Support in UK General Elections [Dataset]. {\bf 2010}. Available online:\\ 
\href{https://www.theguardian.com/news/datablog/2010/may/04/general-election-newspaper-support}{https://www.theguardian.com/news/datablog/2010/may/04/general-election-newspaper-support}
%
\bibitem{foot1}
The unwieldy expression of ${\cal T}(s,1/2)$ is given in \cite{MM2011}.
%
%
\bibitem{foot3}
With 
the  convention $0!!=1$.
%
\bibitem{foot4}
Note a typo in Equation (11) of Ref.~\cite{MM2011} where a factor $1/2$ is missing; compare to Equation \eqref{MET}. The results reported in Fig.3 of Ref.~\cite{MM2011} are however correct.
%
\bibitem{foot5}
The joint Labour/Liberal and Conservative/Liberal support is treated as an influence state $\xi=(1-1)/2=0$.
\end{thebibliography}
\end{document}